\documentclass[aip,preprint,author-year]{revtex4-1}
\usepackage{bm}
\usepackage{amsmath}
\usepackage{amsfonts}
\usepackage{graphicx}
\usepackage{epstopdf}

\begin{document}

\title{Representation and Conservation of Angular Momentum in the 
Born-Oppenheimer Theory of Polyatomic Molecules}
\author{Robert Littlejohn}
\affiliation{Department of Physics, University of California, 
Berkeley, CA, USA}
\author{Jonathan Rawlinson}
\affiliation{School of Mathematics, University of Manchester,
  Manchester, UK}
\author{Joseph Subotnik}
\affiliation{Department of Chemistry, University of Pennsylvania, 
Philadelphia, PA, USA}

\date{\today}

\newcommand{\Amat}{{\mathsf{A}}}
\newcommand{\Aspace}{{\mathcal{A}}}
\newcommand{\avec}{{\mathbf{a}}}
\newcommand{\Avec}{{\mathbf{A}}}
\newcommand{\Bspace}{{\mathcal{B}}}
\newcommand{\bra}[1]{\langle#1\vert}
\newcommand{\braket}[2]{\langle#1\vert#2\rangle}
\newcommand{\Bregion}{{\mathcal{B}}}
\newcommand{\Bregionbar}{{\overline{\mathcal{B}}}}
\newcommand{\BS}{{\mathcal{B}}}
\newcommand{\Bvec}{{\mathbf{B}}}
\newcommand{\bvec}{{\mathbf{b}}}
\newcommand{\CaseIa}{{\hbox{\rm Case~Ia}}}
\newcommand{\CaseIb}{{\hbox{\rm Case~Ib}}}
\newcommand{\CasesIab}{{\hbox{\rm Cases~Iab}}}
\newcommand{\CaseII}{{\hbox{\rm Case~II}}}
\newcommand{\codim}{{\mathop{\textrm{codim}}}}
\newcommand{\Complexes}{\mathbb{C}}
\newcommand{\Cbar}{{\bar C}}
\newcommand{\CS}{{\mathcal{C}}}
\newcommand{\Ctilde}{{\tilde C}}
\newcommand{\Fvec}{{\mathbf{F}}}
\newcommand{\Hbar}{{\bar H}}
\newcommand{\Hspace}{{\mathcal{H}}}
\newcommand{\Integers}{\mathbb{Z}}
\newcommand{\Ivec}{{\mathbf{I}}}
\newcommand{\iquat}{{\mathbf{i}}}
\newcommand{\jquat}{{\mathbf{j}}}
\newcommand{\Jvec}{{\mathbf{J}}}
\newcommand{\ket}[1]{\vert#1\rangle}
\newcommand{\ketbra}[2]{\vert#1\rangle\langle#2\vert}
\newcommand{\kquat}{{\mathbf{k}}}
\newcommand{\Ktilde}{{\tilde K}}
\newcommand{\Kvec}{{\mathbf{K}}}
\newcommand{\Levels}{{I}}
\newcommand{\Lvec}{{\mathbf{L}}}
\newcommand{\matrixelement}[3]{\langle#1\vert#2\vert#3\rangle}
\newcommand{\MOI}{{\mathsf{M}}}
\newcommand{\Ne}{{N_e}}
\newcommand{\Nn}{{N_n}}
\newcommand{\nvechat}{{\hat{\mathbf{n}}}}
\newcommand{\omegavec}{\bm{\omega}}
\newcommand{\Pbar}{{\bar{P}}}
\newcommand{\Proj}{{\mathcal{P}}}
\newcommand{\Pvec}{{\mathbf{P}}}
\newcommand{\Pvecbar}{\bar{\mathbf{P}}}
\newcommand{\pvec}{{\mathbf{p}}}
\newcommand{\Quaternions}{\mathbb{H}}
\newcommand{\Reals}{\mathbb{R}}
\newcommand{\Region}{{\mathcal{R}}}
\newcommand{\Regionbar}{{\overline{\mathcal{R}}}}
\newcommand{\rvec}{{\mathbf{r}}}
\newcommand{\Rvec}{{\mathbf{R}}}
\newcommand{\scalarprod}[2]{\langle #1,#2\rangle}
\newcommand{\sigmavec}{\bm{\sigma}}
\newcommand{\Sspace}{{\mathcal{S}}}
\newcommand{\Svec}{{\mathbf{S}}}
\newcommand{\thetavec}{\bm{\theta}}
\newcommand{\tr}{{\mathop{\textrm{tr}}}}
\newcommand{\Vvec}{{\mathbf{V}}}
\newcommand{\xbar}{{\bar x}}
\newcommand{\xivec}{\bm{\xi}}
\newcommand{\xtilde}{{\tilde x}}
\newcommand{\xvec}{{\mathbf{x}}}
\newcommand{\Xvec}{{\mathbf{X}}}
\newcommand{\Xvecbar}{\bar{\mathbf{X}}}
\newcommand{\Xvectilde}{\tilde{\mathbf{X}}}
\newcommand{\ytilde}{{\tilde y}}
\newcommand{\yvechat}{{\hat{\mathbf{y}}}}
\newcommand{\Yvec}{{\mathbf{Y}}}
\newcommand{\Yvectilde}{\tilde{\mathbf{Y}}}
\newcommand{\ztilde}{{\tilde z}}
\newcommand{\Zvectilde}{\tilde{\mathbf{Z}}}

\begin{abstract}
  This paper concerns the representation of angular momentum operators
  in the Born-Oppenheimer theory of polyatomic molecules and the
  various forms of the associated conservation laws.  Topics addressed
  include the question of whether these conservation laws are exactly
  equivalent or only to some order of the Born-Oppenheimer parameter
  $\kappa=(m/M)^{1/4}$, and what the correlation is between angular
  momentum quantum numbers in the various representations.  These
  questions are addressed both in problems involving a single
  potential energy surface, and those with multiple, strongly coupled
  surfaces; and both in the electrostatic model and those for which
  fine structure and electron spin are important.  The analysis leads
  to an examination of the transformation laws under rotations of the
  electronic Hamiltonian; of the basis states, both adiabatic and
  diabatic, along with their phase conventions; of the potential
  energy matrix; and of the derivative couplings.  These
  transformation laws are placed in the geometrical context of the
  structures in the nuclear configuration space that are induced by
  rotations, which include the rotational orbits or fibers, the
  surfaces upon which the orientation of the molecule changes but not
  its shape; and the section, an initial value surface that cuts
  transversally through the fibers.  Finally, it is suggested that the
  usual Born-Oppenheimer approximation can be replaced by a dressing
  transformation, that is, a sequence of unitary transformations that
  block-diagonalize the Hamiltonian.  When the dressing transformation
  is carried out, we find that the angular momentum operator does not
  change.  This is a part of a system of exact equivalences among
  various representations of angular momentum operators in
  Born-Oppenheimer theory.  Our analysis accommodates large-amplitude
  motions, and is not dependent on small-amplitude expansions about an
  equilibrium position.
\end{abstract}

\maketitle

\section{Introduction}
\label{intro}

This article concerns angular momentum and rotations in the
Born-Oppenheimer theory of polyatomic molecules.  Topics addressed
include the relationship among the various representations of angular
momentum operators and the corresponding conservation laws, as well as
the equivalence among them, and whether that is approximate or exact.
We also address the correlation between angular momentum quantum
numbers in the various representations. We treat both single-surface
and multi-surface problems, and we treat both the simple electrostatic
model for the electronic Hamiltonian as well as models that
incorporate fine structure and electron spin.  We assume the molecule
is isolated, so that the Hamiltonian commutes with both rotations and
time reversal.

This article relies on basic Born-Oppenheimer theory
(\cite{BornOppenheimer27, BornHuang54, BallhausenHansen72, Mead88,
Cederbaum04}) and its application to multisurface problems with
conical intersections (\cite{HerzbergLonguetHiggins63,
LonguetHiggins75, Mead79, Mead83, ThompsonMead85, Yarkony96,
Yarkony97a, Yarkony97b, Gordonetal98, Yarkony01, AdhikariBilling02,
KuppermanAbrol02, Domcke04, Yarkony04a, Yarkony04b,
JasperNangiaZhuTruhlar06, SchuurmanYarkony06, Farajietal12, Matsika12,
Yarkony12, ZhuYarkony16, Gononetal17, Kendrick18, FedorovLevine19,
ChoiVanicek20, Bianetal21, WuSubotnik21}).  An important role is
played by diabatic bases (\cite{Smith69, Baer75, Thomsonetal85,
Pacheretal88, Cederbaumetal89, Pacheretal89, Pacheretal93,
AtchityRuedenberg97, MatsunagaYarkony98, ThielKoppel99, Yarkony99,
Yarkony00, AbrolKuppermann02, Koppel04a, Subotniketal08,
Subotniketal09, RichingsWorth15, ZhuYarkony15, VenghausEisfeld16,
Wangetal19, RichingsHabershon20, LittlejohnRawlinsonSubotnik22}).
Extensive attention is devoted to the derivative couplings, which are
the components of a Mead-Truhlar-Berry vector potential or connection,
part of one of the two gauge theories that appears in molecular
Born-Oppenheimer theory (\cite{MeadTruhlar79, Mead80b, Berry84,
Moodyetal89, Bohmetal91, Bohmetal92, Bohmetal92a, Mead92,
KendrickMead95, Kendricketal02, Child02, Kendrick04,
JuanesMarcosAlthorpeWrede05, Althorpe06, Althorpe12, Wittig12,
ChoiVanicek21}).  Finally, we treat electron dynamics both in the
electrostatic model and also when fine structure and electron spin are
important (\cite{Mead80a, Mead87, Yarkony92, KoizumiSugano95,
SchonKoppel98, MatsikaYarkony01, MatsikaYarkony02a, MatsikaYarkony02b,
WuMiaoSubotnik20, SadovskiiZhilinskii22}).

Our analysis requires a careful treatment of the phase and frame
conventions of the electronic basis states, both adiabatic and
diabatic.  We emphasize that the Born-Oppenheimer treatment is not
well defined without phase conventions, and for this reason we will
spend considerable time describing the geometrical context within
which we work.  That context provides geometrical interpretations of
our procedures and of the resulting formulas, and involves geometrical
structures in the nuclear configuration space.  These include the
rotational orbits or fibers, which are the surfaces upon which the
orientation of the molecule changes but not its shape, and the
section, a kind of initial-value surface that cuts transversally
through the fibers. We use rotation operators for assigning phase and
frame conventions when moving along the rotational fibers, and other
algorithms when moving transversally (along the section).  This
distinction has appeared between the lines in existing literature but
it has not been addressed explicitly, as far as we know, nor has the
geometrical context been brought to light.

In the case of fine-structure models with an odd number of electrons,
the method of assigning phase and frame conventions by means of
rotation operators must be modified, in that an extra spin rotation,
applied to the two elements of a Kramers doublet (what we call
``pseudo-spin''), is necessary to create a single-valued set of basis
states.  This observation seems to be new, and it has an important
impact down the line on the form of the Born-Oppenheimer Hamiltonian
and of the angular momentum.  The basic idea is this.  If a molecule
with an odd number of electrons is subjected to a rigid rotation about
some axis by $360^\circ$, then the spatial part of the electronic
eigenfunctions returns to itself but the spin part suffers a change in
sign.  Therefore assigning phase conventions purely by rotation
operators introduces a discontinuity in the basis states.  The
situation bears some similarity to the $-1$ phase shift that
real electronic eigenfunctions suffer in the electrostatic model when being
continuously carried around a conical intersection.  In that case,
\cite{MeadTruhlar79} suggested introducing a complex phase factor (a
$U(1)$ rotation) to smooth out the discontinuity.   Similarly, in our
case, we suggest introducing an extra spin rotation to remove the
discontinuity encountered when rotating the molecule by $360^\circ$. 

The establishment of phase and frame conventions leads to the
derivation of a number of transformation laws of objects under
rotations, including the electronic Hamiltonian, its matrix elements,
the basis states and the derivative couplings.  We have done this in
several different models of the electronic Hamiltonian.  The resulting
explicit formulas seem to be mostly new, although some of them are quite
clear intuitively and \cite{Yarkony01} has derived some closely
related results in the case of nondegenerate, adiabatic basis
states. Nevertheless, the careful derivation of these results involves
some subtleties, for example, some of the results are only valid under
certain circumstances which we specify.  These transformation laws are
necessary to establish the relationship among the various forms of
angular momentum operators.  We believe our transformation laws for
the derivative couplings are new; they are necessary for showing the
invariance of the Born-Oppenheimer Hamiltonians (in their various
versions) under rotations.

In this article we wish to accommodate large amplitude motions, that
is, ones in which the nuclear displacements are of the order of an
atomic unit or larger.  Such motions occur in isomerization,
photoexcitation, scattering and other processes that are of current
interest.  Therefore we require an understanding of angular momentum
and its conservation that allows such motions and that is not
dependent on small-amplitude expansions about an equilibrium position.

Although the results presented below are most directly related to the
determination of stationary states, many of the lessons derived have
implications for time-dependent quantum mechanical simulations as
well.  There are also semiclassical implications with regards to
surface hopping calculations, as will be described in
Sec.~\ref{conclusions}.  

In this article we do not consider the construction of kinetic energy
operators in internal or shape coordinates, but several of our
results, such as the treatment of phase conventions of electronic
basis states by means of rotation operators, the transformation laws
of the derivative couplings under rotations, and the derivation of the
rotational components of the derivative couplings, are necessary
preliminaries for the construction of such operators when multiple
surfaces, geometric phases, and/or fine structure are important.   The
subject of kinetic energy operators is a large one; we just mention
\cite{WangCarrington00, Kendrick18}, of which the latter reference is
notable for its treatment of multiple potential energy surfaces in
scattering calculations. 

In this article for simplicity we ignore nuclear spin, effectively
treating the nuclei as spinless, distinguishable particles.

We turn now to an outline of the paper.  The purpose of
Sec.~\ref{overviewmain} is to place some of the questions raised by this
paper into a simple context, as a way of making a hopefully painless
introduction to the subject before treating it in all generality.  In
addition, Sec.~\ref{overviewmain} establishes terminology and notation.

Section~\ref{overviewmain} treats a polyatomic molecule in the
electrostatic model for which motion on a single potential energy
surface is a good approximation.  There are two descriptions of the
dynamics, one, the ``molecular,'' which explicitly incorporates the
interactions of all the charged particles, electrons and nuclei; and
the other, the ``Born-Oppenheimer,'' in which the electron dynamics is
incorporated into the potential energy function.  The Hamiltonian in
the molecular representation commutes with the total orbital angular
momentum of the molecule, nuclear plus electronic, what we write as
$\Lvec_n+\Lvec_e$, while the Hamiltonian in the Born-Oppenheimer
representation commutes with the nuclear orbital angular momentum
$\Lvec_n$ alone.  These two conservation laws are presumably
equivalent somehow, but we may ask whether this equivalence is exact
or only valid to some order in the Born-Oppenheimer ordering parameter
$\kappa=(m/M)^{1/4}$ (\cite{BornOppenheimer27}).  In addition there is
the question of the correlation between angular momentum quantum
numbers in the two representations.

Section~\ref{overviewmain} presents an overview of the answers to
these questions, first in the electrostatic model and then
generalizing to models that include fine structure and electron spin.
Finally, Sec.~\ref{overviewmain} presents an overview of the dressing
transformation that block-diagonalizes the Born-Oppenheimer
Hamiltonian, and its effect on angular momentum operators.

After this overview the paper presents a more detailed and rigorous
analysis of questions surrounding angular momentum in Born-Oppenheimer
theory.  Although the problems addressed in Sec.~\ref{overviewmain}
concern motion on a single surface, the rest of the paper, starting
with Sec.~\ref{phaseframeES}, treats multiple, strongly interacting
surfaces; naturally, single-surface problems are covered as a special
case.  Multi-surface problems require that diabatic bases be
incorporated into the discussion of basis states.

Sections~\ref{phaseframeES}--\ref{molecBOreps} deal with the
electrostatic model, presenting results that are later generalized to
various fine-structure models.  Section~\ref{phaseframeES} treats 
phase and frame conventions for the electronic basis states, a
necessary topic since the form of operators in the Born-Oppenheimer
representation depends on these conventions.   The subject of phase
and frame conventions is not as well developed in the literature as it
might be, perhaps because in simple (single-surface, electrostatic)
problems the choice of a phase for the one electronic
eigenstate of interest can be reduced to a $\pm$ sign, which seems
trivial.  It is not, actually, even in this case, but when
degeneracies, multiple surfaces, diabatic bases and spin are taken
into account, phase and frame conventions become a more serious
matter.  

In Sec.~\ref{transfbasisF}, continuing with the electrostatic model,
we consider the transformation properties of the basis states under
rotations.  The basis states can be either adiabatic or diabatic.
This leads to a collection of transformation laws under rotations,
including (\ref{ESHeconjR}) for the electronic Hamiltonian,
(\ref{wbasisxfmrots}) for the basis states and (\ref{Fvecxfmrots}) for
the derivative couplings.  An important consequence of these is
(\ref{Ltotbasiszero}), which says that the electronic basis functions,
with our phase conventions, are invariant under simultaneous rotations
of the electronic and nuclear coordinates.  In the context of
nondegenerate, adiabatic basis states this formula is only a small
step away from the results of \cite{Yarkony01}, but the formula is
notable for its simplicity and in our treatment it incorporates
degeneracies and diabatic bases (and later it is generalized to
include spin). This formula is consequential, being important in the
establishment of the equivalence of various representations of the
angular momentum.  

In Sec.~\ref{molecBOreps} we provide careful definitions of what we
call the ``molecular representation'' and the ``Born-Oppenheimer
representation'' of molecular dynamics, which have been mentioned
previously.   We discuss the invertible mapping between these two and
the corresponding map between linear operators in the two
representations.  Several operators are considered, including the
Hamiltonian and the angular momentum.   As far as the latter is
concerned, we are able to show, using (\ref{Ltotbasiszero}), that
$\Lvec_n+\Lvec_e$ in the molecular representation is exactly
equivalent to $\Lvec_n$ alone in the Born-Oppenheimer representation,
that is, in the Born-Oppenheimer representation, the operator that
looks like the nuclear orbital angular momentum actually includes the
electronic orbital angular momentum.   

In Sec.~\ref{detailsFS} we cover the same territory as in
Sections~\ref{phaseframeES}--\ref{molecBOreps} but with the
fine-structure model for the electronic Hamiltonian.  The cases of
even and odd numbers of electrons are treated separately.  The case of
an even number of electrons is broadly similar to the electrostatic
model, with some notable differences such as the fact that the
nominal, nuclear orbital angular momentum $\Lvec_n$ in the
Born-Oppenheimer representation now includes, from a physical
standpoint, not only the electronic orbital angular momentum but also
the electron spin.  The case of an odd number of electrons presents
many new features, such as the extra spin rotation required in the
phase conventions for the basis states (see
(\ref{wbasisxfmrotsFSodd})) in order to make the basis single-valued.

In Sec.~\ref{dressingxfm} we provide a more detailed treatment of the
dressing transformation that removes off-block-diagonal terms in the
Born-Oppenheimer Hamiltonian. A principal conclusion is that the
dressing transformation does not change the form of angular momentum
operators.  This holds to all orders of the Born-Oppenheimer
perturbation parameter $\kappa$.

Finally, in Sec.~\ref{conclusions}, we present some conclusions.

\section{Overview of Main Results in a Simple Context}
\label{overviewmain}

In this section we discuss the equivalence of different angular
momentum operators in a simple context, in order to highlight the
issues before getting into a detailed or general analysis.  We also
establish some notation. 

\subsection{Nuclear Configuration Space}
\label{nuclearcs}

\begin{figure}
\includegraphics[scale=0.5]{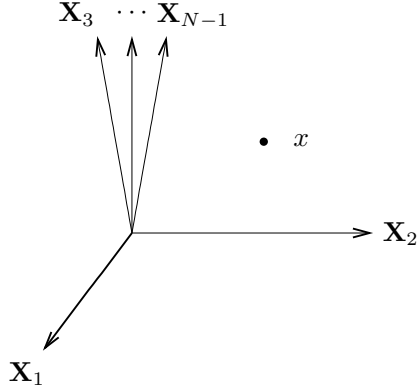}%
\caption{\label{nuclearcsfig}The nuclear configuration space in the
  center-of-mass frame is $\Reals^{3N-3}$, where $N$ is the number of
  nuclei.  This space is indicated schematically by the axes labeled
  $\Xvec_\alpha$, $\alpha=1,\ldots,N-1$.  The notation $x$ stands for
  a point of configuration space, or, equivalently, the coordinates
  $(\Xvec_1,\ldots,\Xvec_{N-1})$ of that point.}
\end{figure}

We assume our molecule has $N\ge3$ nuclei.  To describe the
configuration of the nuclei in the center-of-mass frame we require
$N-1$ translationally invariant vectors, $\Xvec_\alpha$,
$\alpha=1,\ldots,N-1$, the components of which are coordinates on the
nuclear configuration space.  Each component ranges from $-\infty$ to
$+\infty$, so the nuclear configuration space is $\Reals^{3N-3}$.
This is the parameter space for the electronic Hamiltonian; it is
topologically trivial.  For brevity we denote the nuclear coordinates
collectively by $x$, so that
\begin{equation}
  x=(\Xvec_1,\ldots,\Xvec_{N-1}).
  \label{xdef}
\end{equation}
We also use the symbol $x$ to stand geometrically for a point of the
nuclear configuration space, as illustrated in
Fig.~\ref{nuclearcsfig}.

We choose the vectors $\Xvec_\alpha$ to be Jacobi vectors
(\cite{Delves60, AquilantiCavalli86, Gattietal98}), which cause the
nuclear kinetic energy $K_n$ to be diagonal,
\begin{equation}
  K_n = \sum_{\alpha=1}^{N-1} \frac{\Pvec_\alpha^2}{2M_\alpha},
\end{equation}
where $\Pvec_\alpha$, $\alpha=1,\ldots,N-1$ are the momenta
conjugate to Jacobi vectors $\Xvec_\alpha$, and where the
$M_\alpha>0$ are reduced nuclear masses.

\subsection{Molecular and Electronic Hamiltonians}
\label{Hamsmodels}

We write the Hamiltonian for the molecule as
\begin{equation}
  H_{\rm mol} = \sum_{\alpha=1}^{N-1}
  \frac{\Pvec_\alpha^2}{2M_\alpha}
  +H_e(x;\rvec,\pvec,\Svec),
  \label{Hmoldef}
\end{equation}
where $H_e=H_e(x)=H_e(x;\rvec,\pvec,\Svec)$ is the electronic
Hamiltonian and where $\rvec = (\rvec_1,\ldots,\rvec_{N_e})$, $\pvec =
(\pvec_1,\ldots,\pvec_{N_e})$, and $\Svec =
(\Svec_1,\ldots,\Svec_{N_e})$ are the electron positions, momenta and
spins, respectively.  Here $N_e$ is the number of electrons, the
electron positions $\rvec_i$, $i=1,\ldots,N_e$, are measured relative
to the nuclear center of mass, and the electron momenta $\pvec_i$ are
conjugate to the positions $\rvec_i$.  The parametric dependence of
the electronic Hamiltonian on the nuclear configuration $x$ is set off
by a semicolon from the electronic operators $(\rvec,\pvec,\Svec)$
upon which it depends.

The molecular Hamiltonian $H_{\rm mol}$ depends on both nuclear and
electronic operators,
\begin{equation}
  H_{\rm mol}=H_{\rm mol}(\Xvec,\Pvec,\rvec,\pvec,\Svec),
  \label{Hmoldependence}
\end{equation}
where $\Xvec = (\Xvec_1,\ldots,\Xvec_{N-1})$ and $\Pvec =
(\Pvec_1,\ldots,\Pvec_{N-1})$.  The notation $\Xvec$ is essentially
the same as $x$, the only difference being one of emphasis ($\Xvec$
being used for the Jacobi vectors upon which $H_{\rm mol}$ or a wave
function depends, while $x$ stands either for those vectors or a point
of nuclear configuration space).  

\subsection{Models of the Electronic Hamiltonian}
\label{models}

We consider the electronic Hamiltonian in various models.  The most
basic is the electrostatic, for which the electronic Hamiltonian is
\begin{equation}
  H_e(x;\rvec,\pvec) = \sum_{i=1}^{N_e} \frac{\pvec_i^2}{2m_e}
  + \sum_{i,j=1}^{N_e} \frac{\pvec_i\cdot\pvec_j}{2M_n} + 
  V_{\rm Coul}(\Xvec,\rvec),
  \label{HeESdef}
\end{equation}
where $m_e$ is the electron mass, $M_n$ is the total nuclear mass and
where the potential $V_{\rm Coul}$ contains all the Coulomb
interactions among all the particles (electrons and nuclei).  The
second major term is the mass-polarization term, which is due to the
fact that the nuclear center of mass, to which the electron
coordinates $\rvec_i$ are referred, is not fixed in an inertial frame.
In the electrostatic model the electronic Hamiltonian
$H_e=H_e(x;\rvec,\pvec)$ is independent of the electron spin, and so
can be regarded as an operator acting on the space of purely spatial
electronic wave functions $\phi(\rvec)$, that is, with no dependence
on spin quantum numbers $m$.

Other models are obtained by adding fine structure terms to
(\ref{HeESdef}) (\cite{BetheSalpeter57, HowardMoss70, Yarkony92,
  HessMarian00}).  In the resulting fine structure models the
electronic Hamiltonian $H_e=H_e(x;\rvec,\pvec,\Svec)$ does depend on
the spin and the electronic wave function $\phi(\rvec,m)$ depends on
the electron spin quantum numbers,
\begin{equation}
  m=(m_1,\ldots,m_{N_e}),
  \label{mdef}
\end{equation}
where $m_i=\pm 1/2$, $i=1,\ldots,N_e$. There is some latitude in how
relativistic corrections are treated, but in fact the only assumptions
we shall make about the fine structure model are the symmetries of the
electronic Hamiltonian, which apply in all cases.

In the following we use the symbol $\phi$ for a purely electronic wave
function (that is, $\phi(\rvec)$ in the electrostatic model or
$\phi(\rvec,m)$ if electron spin is included); $\psi$ for a purely
nuclear wave function (that is, $\psi(\Xvec)$); and $\Psi$ for a
molecular wave function (that is, $\Psi(\Xvec,\rvec)$ in the
electrostatic model or $\Psi(\Xvec,\rvec,m)$ if electron spin is
included).

\subsection{Two Conservation Laws}
\label{twolaws}

We now pose a set of questions regarding angular momentum conservation
in Born-Oppenheimer theory. For simplicity we do this initially in the
electrostatic model, generalizing later (in Sec.~\ref{answersfs}) to
the fine structure model.  Also, for simplicity, we present our
questions in the context of motion on a single potential energy
surface, generalizing later (starting in Sec.~\ref{phaseframeES}) to
multisurface problems.

Suppose we wish to find energy eigenfunctions for the whole molecule,
that is, wave functions $\Psi(\Xvec,\rvec)$ such  that 
\begin{equation}
  H_{\rm mol}(\Xvec,\Pvec,\rvec,\pvec)\Psi(\Xvec,\rvec) =
  E\,\Psi(\Xvec,\rvec), 
  \label{HPsiEPsi}
\end{equation}
either bound or unbound (see, for example,
\textcite{CafieroAdamowicz04}).  The molecular Hamiltonian in the
electrostatic model (\ref{HeESdef}) commutes with the total orbital
angular momentum of the molecule,
\begin{equation}
    \Lvec = \Lvec_n + \Lvec_e = \sum_{\alpha=1}^{N-1} \Xvec_\alpha
    \times \Pvec_\alpha + \sum_{i=1}^{N_e} \rvec_i\times\pvec_i,
    \label{Lvecdef}
\end{equation}
which we have broken into the nuclear and electronic contributions.
It does not commute with $\Lvec_n$ or $\Lvec_e$ separately.  Therefore
it is possible to organize the energy eigenfunctions (by forming linear
combinations of degenerate energy eigenfunctions, if necessary) to be
also eigenfunctions of the operators $L^2$ and $L_z$.

Instead of (\ref{HPsiEPsi}) one often solves the Born-Oppenheimer version of
the Schr\"odinger equation,
\begin{equation}
  H_{\rm BO}(\Xvec,\Pvec)\psi(\Xvec) = E\,\psi(\Xvec),
  \label{HpsiEpsi}
\end{equation}
where $\psi=\psi(\Xvec)$ is a function of the nuclear coordinates
alone.  The Born-Oppenheimer version of the Hamiltonian is
\begin{equation}
  H_{\rm BO}(\Xvec,\Pvec) = \sum_{\alpha=1}^{N-1} 
  \frac{\Pvec_\alpha^2}{2M_\alpha}
  +\epsilon_k(\Xvec),
  \label{HBOdef}
\end{equation}
where $\epsilon_k(\Xvec)=\epsilon_k(x)$ is the $k$-th eigenvalue of
$H_e(x)$.  This Hamiltonian describes motion on a single potential
energy surface $k$; in practice this is often the ground state.  The
Born-Oppenheimer Hamiltonian (\ref{HBOdef}) is like the molecular one
(\ref{Hmoldef}) except that the electronic Hamiltonian $H_e(x)$ has
been replaced by one of its eigenvalues $\epsilon_k(x)$.

The Born-Oppenheimer Hamiltonian (\ref{HBOdef}) commutes with the
nuclear orbital angular momentum, 
\begin{equation}
  \Lvec_n = \sum_{\alpha=1}^{N-1} \Xvec_\alpha \times \Pvec_\alpha,
  \label{Lvecndef}
\end{equation}
because the electronic eigenvalues are invariant under rotations,
\begin{equation}
  \epsilon_k(x) = \epsilon_k(Rx), \qquad \forall R\in SO(3),
  \label{epsilonkinv}
\end{equation}
where $Rx$ indicates a rigid rotation of  the nuclei about the center
of mass,
\begin{equation}
  Rx = R(\Xvec_1,\ldots,\Xvec_{N-1}) = 
  (R\Xvec_1,\ldots,R\Xvec_{N-1}).
  \label{Rxdef}
\end{equation}
This is because the electronic eigenvalues do not change if the nuclei
are subjected to a rigid rotation, that is, one that changes the
orientation of the nuclei but not their shape.  

Therefore the energy eigenfunctions $\psi(\Xvec)$ of (\ref{HpsiEpsi})
can be organized (by forming linear combinations of degenerate
eigenfunctions, if necessary) to be simultaneous eigenfunctions of
energy, $L_n^2$ and $L_{nz}$, the latter of which refer to the nuclear
orbital angular momentum $\Lvec_n$. 

\subsection{Questions About the Two Conservation Laws}
\label{questions} 

Thus it would appear that the Born-Oppenheimer approximation has
replaced one exact conservation law (that of $\Lvec=\Lvec_n+\Lvec_e$)
with another (that of just $\Lvec_n$).  This leads us to ask, are
these conservation laws just approximate versions of one another, or
are they somehow exactly equivalent?  And how does this come about in
detail?  There is also the question of the physical interpretation of
the solutions of the Born-Oppenheimer equation (\ref{HpsiEpsi}).  If
we find such a solution $\psi(\Xvec)$ that is an eigenfunction of
energy, $L_n^2$ and $L_{nz}$ with quantum numbers $(E,l,m_l)$,
then presumably (as is standard in Born-Oppenheimer theory) the
corresponding solution of (\ref{HPsiEPsi}) will be approximately
\begin{equation}
  \Psi(\Xvec,\rvec)=\psi(\Xvec)\phi_k(\Xvec;\rvec),
  \label{Psipsiphi}
\end{equation}
where $\phi_k(\Xvec;\rvec)$ is the $k$-th energy eigenfunction of the
electronic Hamiltonian $H_e(x)$.   Is this $\Psi$ then an
eigenfunction of $L^2$ and $L_z$ (which refer to the total orbital
angular momentum, $\Lvec=\Lvec_n+\Lvec_e$)?  If so, is it exactly so or
only to some order of the Born-Oppenheimer expansion?  And
are the angular momentum quantum numbers of the solution
$\Psi(\Xvec,\rvec)$ the same as those of $\psi(\Xvec)$, what we have
called $(l,m_l)$, even though  the operators appear to be different?

Finally, how do the answers to these questions change when fine
structure effects are included or when multiple potential energy
surfaces are strongly coupled?

\subsection{Overview of Some Answers in the Electrostatic Model}
\label{someanswers}

It is convenient to introduce ket language for the eigenfunctions
$\phi_k(\Xvec;\rvec)$ of the electronic Hamiltonian $H_e(x)$ in the
electrostatic model.  We denote these eigenkets by $\ket{x;k}$, so
that
\begin{equation}
  H_e(x)\,\ket{x;k} = \epsilon_k(x)\,\ket{x;k},
  \label{HeketEket}
\end{equation}
and so that the relation between the kets and wave functions is given
by
\begin{equation}
  \phi_k(\Xvec;\rvec) = \braket{\rvec}{x;k}.
  \label{phiket}
\end{equation}

We must also address the derivative couplings, which are defined by
\begin{equation}
  \Fvec_{\alpha;kl}(x) = \matrixelement{x;k}{\nabla_\alpha}{x;l},
  \label{Fvecdef}
\end{equation}
where $\nabla_\alpha=\partial/\partial \Xvec_\alpha$.  If we write
simply $\Fvec_\alpha(x)$, we refer to the infinite-dimensional matrix
(really a 3-vector of matrices for each value of $\alpha$) whose
$kl$-th component is $\Fvec_{\alpha;kl}(x)$.  It follows from the
orthonormality of the basis, $\braket{x;k}{x;l}=\delta_{kl}$, that the
matrix $\Fvec_\alpha$ is anti-Hermitian,
\begin{equation}
  \Fvec_{\alpha;kl}=-\Fvec^*_{\alpha;lk}.
  \label{FisantiHerm}
\end{equation}

The questions posed can only be answered relative to the phase conventions
for the electronic eigenstates $\ket{x;k}$.  In the electrostatic
model we will require that the energy eigenfunctions
$\phi_k(\Xvec;\rvec)$ be real, that is, invariant under time reversal.
This reduces the phase convention to a choice of a $\pm$ sign, a
subject that we address more carefully in Sec.~\ref{phaseframeES}.
The reality of the basis functions means that the derivative couplings
$\Fvec_{\alpha;kl}$ are real, which, combined with
(\ref{FisantiHerm}), implies that the matrix $\Fvec_\alpha$ is real
and antisymmetric.  This in turn implies that the derivative couplings
vanish on the diagonal, $\Fvec_{\alpha;kk}=0$, which is why those
couplings do not appear in our single-surface, Born-Oppenheimer
version (\ref{HBOdef}) of the Hamiltonian.

To answer one of our questions in the electrostatic model, it turns
out that the two conservation laws are exactly equivalent to one another.
We can state the matter by recalling that in quantum mechanics,
physical observables are represented by linear operators, but the
linear operator representing a given physical observable depends on
the representation of the quantum states.  If the physical observable
is the total orbital angular momentum of the molecule, nuclear plus
electronic, then, when acting on molecular wave functions
$\Psi(\Xvec,\rvec)$, the linear operator is $\Lvec=\Lvec_n+\Lvec_e$,
as in (\ref{Lvecdef}).  But when acting on wave functions
$\psi(\Xvec)$ in the Born-Oppenheimer representation, the same
physical observable is represented by $\Lvec_n$ alone.  Thus, what
appears to be the nuclear orbital angular momentum, when acting on
$\psi(\Xvec)$, actually includes physically the electronic orbital
angular momentum.  We emphasize that this is exact.

To answer another of our questions, suppose that $\psi(\Xvec)$ is a
solution of (\ref{HpsiEpsi}), a simultaneous eigenfunction of $(H_{\rm
  BO},L_n^2,L_{nz})$ with quantum numbers $(E,l,m_l)$.  Also, let
$\Psi(\Xvec,\rvec)$ be defined by (\ref{Psipsiphi}).  Then it turns
out that $\Psi$ is automatically an eigenfunction of $(L^2,L_z)$ with
the same quantum numbers $(l,m_l)$; and this is exact.  (It is, however,
only approximately an eigenfunction of the molecular Hamiltonian
$H_{\rm mol}$.)

One may wonder how we can claim something is exact when the
Born-Oppenheimer approximation is only an approximation.  The brief
answer is that the Born-Oppenheimer approximation approximates the
Hamiltonian but not the angular momentum.  A more sophisticated point
of view, in which the Born-Oppenheimer approximation is replaced by a
sequence of unitary transformations, will be discussed in
Sec.~\ref{dressedvars} and in greater detail in
Sec.~\ref{dressingxfm}.

\subsection{Answers in the Fine Structure Model}
\label{answersfs}

When fine structure effects are included these results generalize in
interesting ways.  Time reversal plays an important role in this
case (see, for example, \cite{Mead79}).  Time reversal $T$ is an
antiunitary operator that acts on electronic wave functions
$\phi(\rvec,m)$ according to (\ref{TdefFS}) or (\ref{TdefFSabbrev}).
It commutes with the electronic Hamiltonian,
\begin{equation}
  T^\dagger H_e(x) T = H_e(x),
\end{equation}
since our molecule is isolated and not interacting with external
fields.  The properties of time reversal that we will need are
summarized in Appendix~\ref{timerev}. 

In the fine structure model the electronic and molecular Hamiltonians
depend on electron spin $\Svec$, and the molecular Schr\"odinger
equation (\ref{HPsiEPsi}) of the electrostatic model must be replaced
by
\begin{equation}
  H_{\rm mol}(\Xvec,\Pvec,\rvec,\pvec,\Svec)\,
  \Psi(\Xvec,\rvec,m) = E\,\Psi(\Xvec,\rvec,m),
  \label{HPsiEPsiFS}
\end{equation}
where now the molecular wave function $\Psi(\Xvec,\rvec,m)$ depends on
the electron spin quantum numbers $m$.  The molecular Hamiltonian no
longer commutes with $\Lvec$ but it does commute with the total
angular momentum of the molecule,
\begin{equation}
  \Jvec=\Lvec+\Svec=\Lvec_n+\Lvec_e+\Svec
  =\sum_{\alpha=1}^{N-1}\Xvec_\alpha \times \Pvec_\alpha
  +\sum_{i=1}^{N_e} \rvec_i\times\pvec_i
  +\sum_{i=1}^{N_e} \Svec_i,
  \label{Jvecdef}
\end{equation}
that is, including the electron spin.  This is one exact conservation
law in the case of the fine structure model.  Now energy
eigenfunctions of the molecule, $\Psi(\Xvec,\rvec,m)$, solutions of
(\ref{HPsiEPsiFS}), can be organized to be also eigenfunctions of
$J^2$ and $J_z$.

\subsubsection{Even Number of Electrons}
\label{evennumber} 

We treat first the case of an even number of electrons.  We denote the
electronic energy eigenstates in ket language as $\ket{x;k}$, as in
the electrostatic model, so that (\ref{HeketEket}) is still valid, but
the electronic eigenfunctions (\ref{phiket}) must be replaced by
\begin{equation}
  \phi_k(\Xvec;\rvec,m) = \braket{\rvec,m}{x;k},
  \label{phiketm}
\end{equation}
that is, with an $m$-dependence.  We choose the eigenstates
$\ket{x;k}$ to be invariant under time reversal, 
\begin{equation}
  T\ket{x;k}  = \ket{x;k}
  \label{Tinvarianteven}
\end{equation}
(see Sec.~\ref{Tcase+1} for a proof that this can be done).  In the
case of a nondegenerate energy level this is a matter of a phase
convention, which is determined to within a $\pm$ sign, as in the
electrostatic model.  For a single-surface problem, as here, the
relevant level is nondegenerate.

The condition (\ref{Tinvarianteven}) is enough to make the derivative
couplings vanish on the diagonal, as in the electrostatic model, so
the Born-Oppenheimer Hamiltonian is still given by (\ref{HBOdef}),
that is, with no derivative couplings.  The only difference is that
the electronic eigenvalue $\epsilon_k(x)$ now includes fine structure
contributions.  This Hamiltonian still commutes with $\Lvec_n$, the
nominal, nuclear orbital angular momentum (see (\ref{Lvecndef})).
Also, the Born-Oppenheimer wave function is still $\psi(\Xvec)$.

Now the operator representing the total angular momentum of the
molecule, nuclear orbital, electronic orbital, and electronic spin,
when acting on molecular wave functions $\Psi(\Xvec,\rvec,m)$, is
$\Jvec$, given by (\ref{Jvecdef}); while the operator representing the
same physical observable, when acting on Born-Oppenheimer wave
functions $\psi(\Xvec)$, is $\Lvec_n$ alone, given by
(\ref{Lvecndef}); and this is exact.  In other words, $\Lvec_n$, when
acting on Born-Oppenheimer wave functions $\psi(\Xvec)$ in the fine
structure model with $N_e={\rm even}$, includes physically the
electronic angular momentum, both orbital and spin.

In addition, suppose we solve the Born-Oppenheimer version of the
Schr\"odinger equation (\ref{HpsiEpsi}) for a wave function
$\psi(\Xvec)$ that is a simultaneous eigenfunction of energy, $L_n^2$
and $L_{nz}$ with quantum numbers $(E,l,m_l)$, and then we define a
molecular wave function by
\begin{equation}
  \Psi(\Xvec,\rvec,m)=\psi(\Xvec)\,\phi_k(\Xvec;\rvec,m),
  \label{Psipsiphieven}
\end{equation}
a generalization of (\ref{Psipsiphi}), where $k$ is the surface in
question.  Then $\Psi(\Xvec,\rvec,m)$ is exactly an eigenfunction of
$J^2$ and $J_z$ with the same quantum numbers $(l,m_l)$, and
approximately an eigenfunction of energy.  Notice that with an even
number of electrons the quantum number of $J^2$ must be an integer, as
is the quantum number $l$ of the nuclear orbital angular momentum
$L_n^2$ (otherwise our statements would not make sense).

\subsubsection{Odd Number of Electrons}
\label{oddnumber}

In the fine structure model with an odd number of electrons the
electronic energy eigenstates are Kramers doublets (\cite{Messiah66}),
that is, they come in pairs $\ket{x;k\mu}$, $\mu=1,2$, such that
\begin{equation}
  H_e(x)\ket{x;k\mu} = \epsilon_k(x) \ket{x;k\mu},
  \label{Kramerspair}
\end{equation}
in which the energy $\epsilon_k(x)$ does not depend on $\mu$.  We
shall think of a Kramers doublet as corresponding to a single
potential energy surface, so that $k$ labels the surfaces and each
surface corresponds to two degenerate levels. For now for simplicity
we treat the problem of a single surface.  This is realistic, for
example, when fine structure effects are added to a system in a spin
doublet state.

Since $H_e(x)$ commutes with time reversal it is possible to choose
the eigenstates $\ket{x;k\mu}$ so that $T\ket{x;k1}=\ket{x;k2}$,
$T\ket{x;k2}=-\ket{x;k1}$, or, equivalently,
\begin{equation}
  T\ket{x;k\mu} = \sum_\nu \ket{x;k\nu}\,\tau_{\nu\mu},
  \label{basisisquat}
\end{equation}
where $\tau$ is given by (\ref{taudef}), as we shall do.  Such a basis
is said to be {\em quaternionic} (see Sec.~\ref{qbases}).   To say
that the basis is quaternionic only determines that basis to within an
$SU(2)$ transformation (\cite{Mead87}).  We choose the basis so that
it transforms under rotations according to (\ref{wbasisxfmrotsFSodd}). 

The electronic energy eigenfunction corresponding to $\ket{x;k\mu}$
now has a double index,
\begin{equation}
  \phi_{k\mu}(\Xvec;\rvec,m) = \braket{\rvec,m}{x;k\mu},
  \label{phiketT=-1}
\end{equation}
which replaces (\ref{phiketm}).  The Born-Oppenheimer wave function
$\psi_{k\mu}(\Xvec)$ carries the same double index, and the
molecular wave function is given by
\begin{equation}
  \Psi(\Xvec,\rvec,m) = \sum_\mu \psi_{k\mu}(\Xvec)\,
  \phi_{k\mu}(\Xvec;\rvec,m),
  \label{PsipsiphiFS-1}
\end{equation}
that is, with a sum over $\mu$.  There is no sum on $k$ because we are
working on a single surface. 

In the fine structure model with $N_e={\rm odd}$ the Born-Oppenheimer
Hamiltonian contains derivative couplings, even for a single surface,
because there is always more than one level (two, for a single
surface).  Now the derivative couplings also carry doubled indices,
\begin{equation}
  \Fvec_{\alpha;k\mu,l\nu}(x) = 
  \matrixelement{x;k\mu}{\nabla_\alpha}{x;l\nu},
  \label{FvecdefT=-1}
\end{equation}
which we can break up into minor, $2\times 2$ matrices as in
Sec.~\ref{qbases}.   That is, in the context of an odd number of
electrons, when we write $\Fvec_{\alpha;kl}$ we mean the minor
($2\times2$) matrix whose $(\mu\nu)$ component is
$\Fvec_{\alpha;k\mu,l\nu}$.  Because of the orthonormality relations,
$\braket{x;k\mu}{x;l\nu}=\delta_{kl}\,\delta_{\mu\nu}$, the derivative
couplings satisfy
\begin{equation}
  \Fvec_{\alpha;k\mu,l\nu}=-\Fvec^*_{\alpha;l\nu;k\mu},
  \label{FisantiHermT=-1}
\end{equation}
a generalization of (\ref{FisantiHerm}), which in the language of
minor matrices becomes 
\begin{equation}
  \Fvec_{\alpha;kl} = -(\Fvec_{\alpha;lk})^\dagger.
  \label{FisantiHermT=-1v2}
\end{equation}

As for the Born-Oppenheimer Hamiltonian, a standard way of deriving it
is to project the molecular Hamiltonian onto a subspace of chosen
energy levels (\textcite{Yarkony96, Cederbaum04}), which in this case
is the subspace spanned by $\ket{x;k\mu}$ for fixed $k$ and $\mu=1,2$.
Doing this we obtain the Born-Oppenheimer version of the Schr\"odinger
equation,
\begin{eqnarray}
  &\displaystyle \sum_{\nu=1}^2 \Bigl[\sum_{\alpha=1}^{N-1}\frac{1}{2M_\alpha}
    \bigl( \Pvec_\alpha^2 \,\delta_{\mu\nu}
    -2i\hbar\,\Fvec_{\alpha;k\mu,k\nu}\cdot\Pvec_\alpha
    -\hbar^2\,G_{\alpha;k\mu,k\nu}\bigr)\nonumber\\
    &+\epsilon_k(\Xvec)\,\delta_{\mu\nu}\Bigr]
    \psi_{k\nu}(\Xvec)
	=E\,\psi_{k\mu}(\Xvec),
    \label{HpsiEpsiFST=-1}
\end{eqnarray}
which replaces (\ref{HpsiEpsi}) and (\ref{HBOdef}) in the
electrostatic model.  Here we define
\begin{equation}
  G_{\alpha;k\mu,l\nu} = \matrixelement{x;k\mu}{\nabla^2_\alpha}
  {x;l\nu},
  \label{Galphakmulnudef}
\end{equation}
which gives us minor matrices $G_{\alpha;kl}$ (and note that only the
diagonal elements $k=l$ of $\Fvec$ and $G$ appear in the Hamiltonian
in (\ref{HpsiEpsiFST=-1})).  This notation is close to that used by
\cite{Cederbaum04} in the electrostatic model.

Since the operators $\nabla_\alpha$ and $\nabla_\alpha^2$ commute with
$T$, the minor matrices $\Fvec_{\alpha;kl}$ and $G_{\alpha;kl}$ are
quaternions (see (\ref{Aisquatproof}); $\nabla_\alpha$ and
$\nabla^2_\alpha$ are not linear operators in the usual sense but the
proof goes through just the same).  In the language of quaternions
(\ref{FisantiHermT=-1v2}) becomes
\begin{equation}
  \Fvec_{\alpha;kl} = -\overline{\Fvec_{\alpha;lk}}.
  \label{FisantiHermT=-1v3}
\end{equation}
For our single-surface problem we need only the diagonal elements
$(k=l)$ of the derivative couplings, which satisfy $\Fvec_{\alpha;kk}
=-\overline{\Fvec_{\alpha;kk}}$, that is, they are quaternions whose
real part (the $a$-part of (\ref{qdef})) vanishes.  We see that the
derivative couplings for a single surface in the case of an odd number
of electrons can be written as a purely imaginary, linear combination
of the Pauli matrices (see also \cite{Mead87}).

We write the $i$-th component of $\Fvec_\alpha$, for $i=1,2,3$, as
$F_{i\alpha}$, and then define coefficients $A_{ji\alpha;kk}$ by
\begin{equation}
  F_{i\alpha;kk} = -\frac{i}{2} \sum_{j=1}^3\sigma_j\,A_{ji\alpha;kk},
  \label{Ajialphadef}
\end{equation}
where $\sigma_j$ are the Pauli matrices.  In this formula we have
split off a factor of $-i$ as in the $\bvec$-part of (\ref{qdef}),
which makes the coefficients $A_{ji\alpha;kk}$ real, and introduced a
factor of $1/2$ for convenience.

The Born-Oppenheimer wave function $\psi_{k\mu}$ for fixed $k$ and
$\mu=1,2$ looks like the wave function of a pseudo-particle with
spin $1/2$, moving on a multidimensional potential energy
surface given by $\epsilon_k(x)$. We define the pseudo-spin operator,
\begin{equation}
  \Kvec=\frac{\hbar}{2}\sigmavec,
  \label{Kvecdef}
\end{equation}
so that the Born-Oppenheimer Hamiltonian can be written as
\begin{equation}
  H_{\rm BO}=\sum_{\alpha=1}^{N-1}\frac{1}{2M_\alpha}
  (\Pvec_\alpha^2 -2\Kvec\cdot\Amat_{\alpha;kk}\cdot\Pvec_\alpha
  -\hbar^2 G_{\alpha;kk})+\epsilon_k(\Xvec),
  \label{HBOFST=-1}
\end{equation}
where $\Amat_{\alpha;kk}$ is a real, $3\times3$ tensor whose
$ji$-component is $A_{ji\alpha;kk}$.  This is written in the style
common with the Pauli equation, in which all operators are understood
to be $2\times2$ matrices, and scalars are understood to be multiplied
by the unit matrix.

Now we can state the main result.   The total, physical angular
momentum of the molecule, nuclear orbital, electronic orbital plus
electronic spin, is represented by the operator $\Jvec$ (see
(\ref{Jvecdef})) when acting on molecular wave functions
$\Psi(\Xvec,\rvec,m)$; and it is represented by $\Ivec$, defined by
\begin{equation}
  \Ivec=\Lvec_n+\Kvec,
  \label{Ivecdef}
\end{equation}
when acting on Born-Oppenheimer wave functions $\psi_{k\mu}(\Xvec)$.
The latter is the nominal, nuclear orbital angular momentum plus the
pseudo-spin; and this result is exact.  

The angular momentum $\Ivec$ commutes with the Born-Oppenheimer
Hamiltonian (\ref{HBOFST=-1}).  We defer the proof of this since it
involves the transformation properties of the derivative couplings
under rotations, a topic that we take up in Sec.~\ref{HamFSodd}.
But it means that when we solve the Born-Oppenheimer version of the
Schr\"odinger equation in the case of an odd number of electrons, we
can organize the energy eigenfunctions to be also eigenfunctions of
the operators $I^2$ and $I_z$, with (say) quantum numbers $(i,m_i)$.
If we then define a molecular wave function by (\ref{PsipsiphiFS-1}),
it turns out be an exact eigenfunction of $J^2$ and $J_z$ with the
same quantum numbers $(i,m_i)$.  Notice that both $\Ivec$ and $\Jvec$
are half-integral (otherwise our statements would not make sense).

\subsection{Dressed Variables}
\label{dressedvars}

In this article we are drawing a distinction between what we are
calling the molecular representation of wave functions and the
Born-Oppenheimer representation (for example, in the case of the
electrostatic model, this means wave functions $\Psi(\Xvec,\rvec)$ and
$\psi(\Xvec)$, respectively).  \cite{Cederbaum04} has referred what we
call the Born-Oppenheimer representation as a ``dressed''
representation.  The notion of dressing has been used in a different
sense by \cite{MartinazzoBurghardt22}, in connection with electronic
friction.  We prefer to reserve the term ``dressed'' for
representations that are obtained from the Born-Oppenheimer
representation by a sequence of unitary transformations, the purpose
of which is to remove the off-diagonal terms in the molecular
Hamiltonian.  These unitary transformations take the place of what is
usually called the ``Born-Oppenheimer approximation,'' which means
simply neglecting those terms on the grounds that they are small.

If we take the point of view that the Born-Oppenheimer version of
the Schr\"odinger equation (for example, (\ref{HpsiEpsi}) and
(\ref{HBOdef}) or (\ref{HpsiEpsiFST=-1})) is obtained, not by throwing
away terms that couple the various levels, but by transforming them
away, then the operators that appear in the Born-Oppenheimer
Hamiltonian must be interpreted as dressed variables.  In particular,
the operator $\Xvec_\alpha$ no longer represents a Jacobi vector of
the nuclei, but rather it has higher order corrections in the
Born-Oppenheimer parameter $\kappa$.  Another consequence is that
$|\psi(\Xvec)|^2$ no longer represents the probability distribution of
the nuclei in nuclear configuration space, not exactly, anyway, since
there are higher order corrections in $\kappa$.  Similar statements
can be made about the electric current.  Such distinctions can be
important in the analysis of matrix elements involved in radiative
transitions (see, for example, \cite{MeadMoscowitz67, Scherreretal15,
  SchauppEngel20}).

These unitary transformations, which diagonalize the molecular
Hamiltonian leaving Born-Oppenheimer Hamiltonians for the various
surfaces on the diagonal, then create an infinite sequence of dressed
representations, as the off-diagonal coupling terms are removed order
by order.  The question then arises as to what happens to our exact
representations of angular momentum operators as the variables are
dressed. 

The answer is that nothing happens to them, for example, in the
electrostatic model the total orbital angular momentum of the
molecule, nuclear plus electronic, is represented in each of these
dressed representations by the same linear operator $\Lvec_n$ given by
(\ref{Lvecndef}), and this is exact.  This is because the generators
of the unitary transformations that carry out the diagonalization are
scalars, and commute with $\Lvec_n$, and therefore so do the unitary
transformations themselves.  The dressing of $\Lvec_n$ just reproduces
$\Lvec_n$.  Similar statements hold in the fine structure models.

This concludes the overview of our main results.  We turn now to a
more detailed development.

\section{Phase and Frame Conventions of Electronic Basis States}
\label{phaseframeES}

Energy eigenstates are only determined to within a phase (when
nondegenerate) or to within an orthonormal frame in the eigenspace
(when degenerate), and these must be carefully specified as our main
results depend on them.  Notice that a frame in a one-dimensional
space is the same as a phase, so phase and frame conventions are the
same thing.  Similar issues apply to other basis states (diabatic,
etc.) that are not energy eigenstates.  In this section we explain how
phase and frame conventions are related to the geometry of orientation
and shape in configuration space.  We work in the electrostatic model,
deferring fine structure effects until Sec.~\ref{detailsFS}.  For
generality we treat multisurface problems, which include single
surface problems as a special case.

\subsection{Electronic Rotation Operators and  the Electronic Hamiltonian}
\label{ESrotationops}

See Appendix~\ref{rotations} for basic facts about the rotation groups
$SO(3)$ and $SU(2)$.  Electronic orbital rotation operators, denoted
$U_{eo}(R)$, are parameterized by rotations $R\in SO(3)$ and are
defined by their action on electronic wave functions,
\begin{equation}
	\bigl(U_{eo}(R)\phi\bigr)(\rvec) = \phi\bigl(R^{-1}\rvec\bigr),
	\label{Ueoaction}
\end{equation}
where $R^{-1}\rvec$ means $(R^{-1}\rvec_1,\ldots,R^{-1}\rvec_{N_e})$.
We also write $U_{eo}(\nvechat,\theta) =U_{eo}\bigl(
R(\nvechat,\theta)\bigr)$ for these operators in axis-angle form.
They are given in terms of their generators by
\begin{equation}
  U_{eo}(\nvechat,\theta) = \exp\left(-\frac{i}{\hbar}\theta
    \nvechat\cdot\Lvec_e\right).
  \label{Ueodef}
\end{equation}
It follows from (\ref{Ueoaction}) that the operators $U_{eo}(R)$
form a representation of $SO(3)$,
\begin{equation}
  U_{eo}(R_1)U_{eo}(R_2) = U_{eo}(R_1R_2).
  \label{Ueorepn}
\end{equation}

The electrostatic, electronic Hamiltonian $H_e(x)=H_e(x;\rvec,\pvec)$,
given by (\ref{HeESdef}), is a function of the dot products of the
vectors $\Xvec_\alpha$, $\rvec_i$ and $\pvec_i$, and is therefore
invariant if each of these is rotated by the same rotation,
\begin{equation}
  H_e(x;\rvec,\pvec) = H_e(Rx;R\xvec,R\pvec), \qquad \forall R\in SO(3),
  \label{ESH_einvR}
\end{equation}
where $Rx$ is given by (\ref{Rxdef}) and where
\begin{eqnarray}
  R\rvec &=& R(\rvec_1,\ldots,\rvec_{N_e})
  =(R\rvec_1,\ldots,R\rvec_{N_e}),\\
  R\pvec &=& R(\pvec_1,\ldots,\pvec_{N_e})
  =(R\pvec_1,\ldots,R\pvec_{N_e}),
  \label{RonrpX}
\end{eqnarray}
Equation~(\ref{ESH_einvR}) is a statement about the functional form of
the electronic Hamiltonian in the electrostatic model.

On the other hand, the electronic position and momentum operators
transform under conjugation by rotations according to
\begin{equation}
  U_{eo}(R)\, \rvec_i \,U_{eo}(R)^\dagger = R^{-1} \rvec_i,
  \qquad
  U_{eo}(R)\, \pvec_i \,U_{eo}(R)^\dagger = R^{-1} \pvec_i,
  \label{rpvectoropsj}
\end{equation}
which is a statement that $\rvec_i$ and $\pvec_i$ are vector operators
(\cite{Messiah66, Varshalovichetal88}).  Therefore
\begin{equation}
  U_{eo}(R)\, H_e(x;\rvec,\pvec)\, U_{eo}(R)^\dagger =
  H_e\bigl(x;R^{-1}\rvec,R^{-1}\pvec\bigr)
  =H_e(Rx;\rvec,\pvec),
  \label{ESHeconjR0}
\end{equation}
where in the first step the conjugation does nothing to the parameters
$x$ which are just $c$-numbers as far as the rotation operators
$U_{eo}(R)$ are concerned, and where in the second step we have
multiplied all arguments by $R$, which according to (\ref{ESH_einvR})
does not change the answer.  Now simplifying the notation by making
the replacement $H_e(x;\rvec,\pvec) \to H_e(x)$, we can summarize the
result by writing
\begin{equation}
  U_{eo}(R)\,H_e(x)\,U_{eo}(R)^\dagger = H_e(Rx).
  \label{ESHeconjR}
\end{equation}
This is the transformation law for the electrostatic, electronic
Hamiltonian under proper rotations.

\subsection{Rotational Orbits and Fibers}
\label{rotorbits}

\begin{figure}
\includegraphics[scale=0.5]{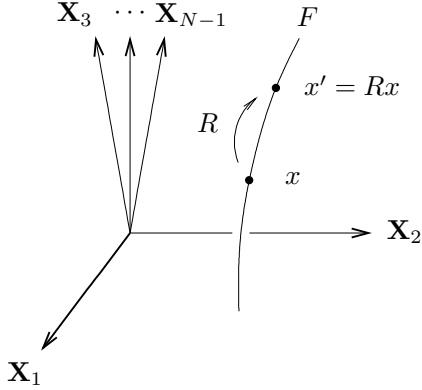}%
\caption{\label{action}A proper rotation $R$ acts on a point $x$ of
  nuclear configuration space and maps it to $x'=Rx$.  The set $F$ of all
  such points $x'$ swept out as $R$ runs over $SO(3)$ is the orbit of
  $x$ under the action of $SO(3)$.  If $x$ is noncollinear, $F$ is a
  fiber in the rotational fiber bundle.}
\end{figure}

The formula (\ref{ESHeconjR}) has a geometrical interpretation in the
nuclear configuration space, which is illustrated in
Fig.~\ref{action}.  Given a configuration $x$ as illustrated, the
rotated configuration $x'=Rx$ is one with the same shape as $x$ but a
different orientation.  Equation~(\ref{ESHeconjR}) relates  the
electronic Hamiltonians at the original point $x$ and the rotated
point $x'$.

Figure~\ref{action} calls attention to the surface $F$, which is the
set swept out by $x'=Rx$ for fixed $x$ as $R$ runs over $SO(3)$.  This is
otherwise the {\it orbit} of $x$ under the action of $SO(3)$ on the
nuclear configuration space.  It is the set of all configurations of
the same shape as $x$ but different orientations.

\begin{figure}
\includegraphics[scale=0.5]{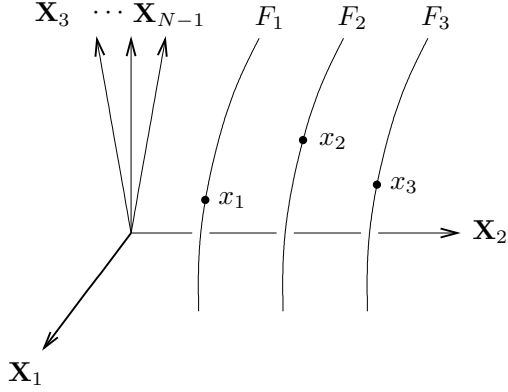}%
\caption{\label{orbits}The action of rotations $R\in SO(3)$ on nuclear
configuration space decomposes that space into a disjoint set of
orbits.}
\end{figure}  

Other configurations of different shapes have their own orbits, as
illustrated in Fig.~\ref{orbits}. Two configurations $x$, $x'$ belong
to the same orbit if and only if there exists $R\in SO(3)$ such that
$x'=Rx$.  Configurations such as $x_1$, $x_2$ and $x_3$ in
Fig.~\ref{orbits}, which do not belong to the same orbit, have
different shapes.  The action of $SO(3)$ decomposes nuclear
configuration space $\Reals^{3N-3}$ into a disjoint set of orbits,
each with its own shape.

The orbits $F$ or $F_i$ illustrated in Figs.~\ref{action} and
\ref{orbits} are drawn as if they were one-dimensional, but actually
their dimensionality is either 0, 2 or 3.  If the configuration is the
$N$-body collision, in which all nuclei are on top of one another,
then rotations do nothing to the configuration and the orbit consists
of a single point, a zero-dimensional set.  If the configuration is
collinear but not the $N$-body collision, then the orientation is
specified by a unit vector along the line of collinearity and the
orbit is diffeomorphic (see Appendix~\ref{rotations}) to the 2-sphere
$S^2$, the space of such unit vectors.  Finally, if the configuration
is noncollinear, then the orbit is diffeomorphic to $SO(3)$ which in
turn is diffeomorphic to $\Reals P^3$ (see
Appendix~\ref{rotations}). This is because two noncollinear
configurations of the same shape are related by a unique $R\in SO(3)$.

In polyatomic molecules most configurations are noncollinear (they
form a subset of full dimensionality, that is, $3N-3$), while the
collinear configurations (and the $N$-body collision, which we count
as collinear) form a subset of measure zero.  In this article we
ignore the collinear configurations, and work only in regions where
all configurations are noncollinear.  We do this for simplicity; the
collinear configurations are the setting for the Renner-Teller effect
(with an extensive literature, including \cite{PericPeyerimhoff02,
  Jungen19, Gamalloetal21}), which is outside the scope of this
article.  For the same reason we restrict consideration to polyatomic
molecules; all diatomics are collinear, and in some ways are more
complicated than polyatomics.

Thus, in the noncollinear subset of nuclear configuration space all
orbits are 3-dimensional.  This subset is decomposed by rotations into
a $(3N-6)$-parameter family of 3-dimensional orbits, each of which is
diffeomorphic to $SO(3)$.  This gives this subset the structure of a
principal fiber bundle (\cite{Nakahara03, Frankel97}), in which the
fibers are the rotational orbits.  For the noncollinear shapes, the
fibers and the rotational orbits are the same thing; in this article
we shall usually refer to them as the ``rotational fibers.''  

\subsection{The Strongly Coupled Subspace}
\label{coupledss}

A pair of adjacent electronic energy levels is considered strongly
coupled if the corresponding energy eigenvalues are degenerate or
nearly degenerate.  This statement is made more quantitative in
Sec.~\ref{dressingxfm}.  We consider a region of nuclear configuration
space in which a chosen subset $\Levels$ of $N_l$ adjacent energy
levels,
\begin{equation}
  \Levels=\{k_0,\ldots,k_0+N_l-1\},
  \label{Adef}
\end{equation}
is not strongly coupled to levels outside of the set $\Levels$, that is,
level $k_0$ is not strongly coupled to level $k_0-1$ and level
$k_0+N_l-1$ is not strongly coupled to level $k_0+N_l$. (Note that if
$k_0$ is the ground state, then there is no level $k_0-1$.) Since the
energy levels are a function of the nuclear configuration $x$, these
conditions can normally hold only over some region of the nuclear
configuration space.

Levels within the set $\Levels$, however, are allowed to be strongly coupled
among themselves, at least somewhere in the region in question.  These
are the conditions that allow a theoretical treatment of the levels
$k\in\Levels$ in isolation from the levels $k\notin\Levels$.  In other words,
degeneracies or near degeneracies that cross the boundaries of $\Levels$ are
not allowed, while internal degeneracies or near degeneracies, those
that take place among the levels $k\in\Levels$, are allowed.

As a special case, in a single-surface problem, $N_l=1$ and $\Levels$
contains the single level $k_0$.  Then internal degeneracies do not
occur, and the condition on the region is that $k_0$ is not degenerate
or nearly degenerate with levels $k_0\pm1$.

These restrictions on the region may cause it to be topologically
nontrivial, either not simply connected or noncontractible, which has
implications for the existence of smooth fields of frames.

We define the strongly coupled subspace $\Sspace(x)$ as the subspace
of the electronic Hilbert space spanned by energy eigenstates for
$k\in\Levels$, and we denote the complementary, orthogonal subspace by
$\Sspace^\perp(x)$.  

\subsection{The Adiabatic Basis}
\label{adiabaticbasis}

It is customary to call the energy eigenbasis the ``adiabatic basis''
but for reasons discussed in \cite{LittlejohnRawlinsonSubotnik22} we
prefer not to work with energy eigenstates for $k\notin\Levels$.  Therefore
we define a set of basis states $\ket{ax;k}$ that are energy eigenstates when
$k\in\Levels$, while for $k\notin\Levels$ we simply require the states
$\ket{ax;k}$ to form a discrete, orthonormal set that spans
$\Sspace^\perp(x)$.  We will call the set $\{\ket{ax;k}\}$ for all $k$
the ``adiabatic basis'' (hence the $a$), but we must remember that
these are energy eigenstates only for $k\in\Levels$.

In addition, we require the basis states to be invariant under
time-reversal, $T\ket{ax;k}=\ket{ax;k}$.  In the electrostatic model,
this just means that the wave functions corresponding to $\ket{ax;k}$
are real (see (\ref{TdefES})).  The energy eigenspaces for $k\in\Levels$
are $T$-invariant, as is $\Sspace(x)$, the sum of such spaces, as is
$\Sspace^\perp(x)$, the orthogonal space (see Appendix~\ref{timerev}).
According to Sec.~\ref{Tcase+1}, this guarantees the existence of a
$T$-invariant basis $\ket{ax;k}$.  A simpler argument that works in
the electrostatic model is that a real Hamiltonian has real
eigenfunctions, but the argument as given generalizes to cases
involving spin.

The freedom in phase and frame conventions that remains after
time-reversal invariance is imposed is the following.  For $k\in\Levels$,
nondegenerate energy eigenstates $\ket{ax;k}$ are determined to within
a $\pm$ sign; for $n$-fold degeneracies inside the strongly coupled
subspace the choices are labeled by elements of the orthogonal group
$O(n)$; and for $k\notin\Levels$ the choices are labeled by the
infinite-dimensional orthogonal group.  (Note that in the case $n=1$,
that is, the nondegenerate case, the group $O(1)$ consists of two
matrices, $(1)$ and $(-1)$, containing the relevant $\pm$ sign.)

Some such choice can be made at each point $x$ of nuclear
configuration space, and is implied in the use of the notation
$\ket{ax;k}$.  We must ask whether these choices can be made in a
smooth manner as $x$ is varied, because discontinuities in the basis
produce divergences in the derivative couplings, which appear in the
Hamiltonian.  In addition, perturbation theory generates derivatives
of the basis states that must be defined and that must have magnitudes
that are under control.  This question can be decomposed into what
happens when we vary the just the orientation, holding the shape
fixed, and what happens when we vary the shape as well.

\subsection{How Phase Conventions Depend on Orientation}
\label{phaseorient}

Let $x_0$ be a noncollinear configuration and let us choose definite
phase and frame conventions for the basis vectors $\ket{ax_0;k}$,
which we assume are $T$-invariant. Thus we have the adiabatic basis
$\ket{ax_0;k}$ at the one point $x_0$.

Now let $x=Rx_0$ for some $R\in SO(3)$, so that $x$ has the same shape
but a different orientation from $x_0$, and define
$\ket{b}=U_{eo}(R)\ket{ax_0;k}$.  We note first that since
time reversal commutes with rotations, $T\ket{b}=\ket{b}$.  Next, if
$k\in\Levels$, then
\begin{eqnarray}
  H_e(x)\ket{b} &=& U_{eo}(R)\,H_e(x_0)\, U_{eo}(R)^\dagger\,
  U_{eo}(R) \,\ket{ax_0;k}\nonumber\\
  &=& U_{eo}(R) \,\epsilon_k(x_0)\,
  \ket{ax_0;k} = \epsilon_k(x_0)\,\ket{b},
\end{eqnarray}
where in the first step we use (\ref{ESHeconjR}).  Thus, $U_{eo}(R)$
maps energy eigenstates at $x_0$ into those at $x=Rx_0$, without
changing the eigenvalues.  More generally, since $U_{eo}(R)$ is
unitary, it maps orthonormal eigenbases inside eigenspaces (degenerate
or not) at $x_0$ into other such bases at $x$.  The fact that the
eigenvalues do not change means that they are invariant under
rotations, as already noted (see (\ref{epsilonkinv})).  As for the
vectors $k\notin\Levels$, $U_{eo}(R)$ maps the orthonormal, $T$-invariant
frame in $\Sspace^\perp(x_0)$ into another such frame in
$\Sspace^\perp(x)$.

We can think of $x_0$ as an initial condition on the fiber passing
through $x_0$.  Since $x_0$ is noncollinear, if $x$ lies on this fiber
then there is a unique $R\in SO(3)$ such that $x=Rx_0$, and point $x$
can be parameterized by $R$.  This allows us to define basis vectors
at $x$, including their phase conventions, by
\begin{equation}
  \ket{ax;k} = U_{eo}(R)\,\ket{ax_0;k},
  \label{adiabphaseconv}
\end{equation}
where $x=Rx_0$. The arbitrarily chosen phase conventions at $x_0$ are
propagated along the rotational fiber by means of rotation operators.

This approach does not work for collinear shapes, for which there is
more than one $R$ that maps a configuration $x_0$ into another one $x$
of the same shape.  Phase conventions for collinear shapes are a more
complicated matter, which we do not cover in this article.

There are other ways of extending phase conventions from a given
point.  In the nondegenerate case a $T$-invariant energy eigenstate
$\ket{ax;k}$ for $k\in\Levels$ is determined to within a $\pm$ sign, a
discrete choice, and the obvious way to extend the phase convention
away from a given point $x_0$ is to demand continuity of the wave
function as $x$ is continuously varied along a path.  We will call
this method, ``extension by continuity.''  It leads to the question of
whether the result depends on the path.  The answer can be developed
in terms of the fundamental group of the region in question (also
called the first homotopy group), as explained by
\cite{JuanesMarcosAlthorpeWrede05, Althorpe06, Althorpe12}.  If the
region is simply connected then the fundamental group is trivial and
extension by continuity gives a unique answer that is a smooth
function of the final position $x$.  If it is not simply connected
then the result may be path-dependent.  (A region is simply connected
if all loops can be smoothly contracted to a point.)

In fact, $SO(3)$ is not simply connected (its fundamental group is
$\Integers_2$), so there is a question as to whether extension of
phase conventions by continuity gives an answer over a rotational
fiber that is path-dependent.  This question is answered, however, by
our formula (\ref{adiabphaseconv}).  Along a noncollinear rotational
fiber, for fixed $x_0$ and variable $x=Rx_0$, $R$ is a smooth function
of $x$, so (\ref{adiabphaseconv}) gives phase conventions over the
fiber that are single-valued, smooth functions of $x$.  Since they are
smooth, they are the same as the conventions obtained by extension by
continuity; and since they are single-valued, the latter method is
path-independent.  

The method of extension by continuity only works when the choices are
discrete, but for degenerate eigenvalues $k\in\Levels$ the choices are
continuous, as are the choices for the orthogonal space
$\Sspace^\perp$ (which is
infinite-dimensional). Equation~(\ref{adiabphaseconv}) works in all
cases and gives phase and frame conventions that are smooth and
single-valued over a rotational fiber.

\subsection{How Phase Conventions Depend on Shape}
\label{phaseshape}

\begin{figure}
\includegraphics[scale=0.5]{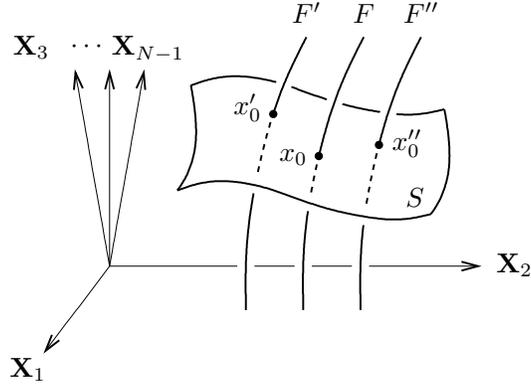}%
\caption{\label{section}Initial points $x_0$ over a family of
  rotational fibers sweep out a surface $S$, a section of  the
  rotational fiber bundle.}
\end{figure}

Now we extend the phase conventions for the basis states to a region
in which both shape and orientation are variable.  We choose some
region of nuclear configuration space consisting of a family of
rotational fibers, as illustrated in Fig.~\ref{section}, and we choose
initial points $x_0$ on each fiber.  If the assignment of the initial
points $x_0$ is made in a smooth manner, these points sweep out a
smooth surface called a {\it section} of the fiber bundle, which is
denoted $S$ in Fig.~\ref{section}.  If we can make a smooth assignment
of phase and frame conventions for our basis along $S$, then we can
use (\ref{adiabphaseconv}) to smoothly extend those conventions along
the rotational fibers.

The section should have dimensionality $3N-6$ so that, taken with the
3-dimensional fibers, it covers a region of nuclear configuration
space of full dimensionality $3N-3$.  As we move along $S$ the shape
of the molecule changes, so coordinates on $S$ can be taken to be
shape coordinates.  These are collections of $3N-6$ rotationally
invariant functions of the Jacobi vectors $\Xvec_\alpha$.  In
practice, bond lengths and angles are common choices for shape
coordinates.  We require that the section be transverse (not tangent)
to the rotational fibers, so that first order displacements along $S$
produce first order changes in shape (this condition makes certain
Jacobian matrices well behaved).

Let us now arbitrarily choose phase and frame conventions for the
basis states $\ket{ax_0;k}$ at one point $x_0$ on the section, as in
Fig.~\ref{section}, and ask if those conventions can be extended in
a smooth manner to neighboring points on the section such as $x'_0$
and $x''_0$ in the figure.

For simplicity let us begin with a single surface problem, for which
$\Levels$ contains the single, nondegenerate level $k_0$.  Then the choice
of phase convention for $\ket{ax;k_0}$ is that of a $\pm$ sign, a
discrete choice, and the method of extension by continuity can be
applied.  This shows that it is possible to make a smooth assignment
of phase conventions for the adiabatic basis vector $k=k_0$ in simply
connected regions of $S$.  

In some cases, however, the region of interest is not simply
connected, as when it encircles a conical intersection.  Then it turns
out that the nondegenerate energy eigenstate $\ket{ax;k_0}$, when carried
continuously as $x$ encircles the conical intersection, undergoes a
sign change on returning to its initial point.  In this case one can
break the region into subregions that are simply connected, with
transition rules in the overlaps to connect them together.  In
practice an equivalent method is preferred, one that employs a single
region with an enforced discontinuity (a change in sign) along a
boundary surface.  This situation is well known and well understood
(\cite{HerzbergLonguetHiggins63, LonguetHiggins75, MeadTruhlar79,
  Mead92, Yarkony96, KuppermanAbrol02, Althorpe06}), but the usual
discussions pay no attention to the geometry of the rotational orbits
and the section, or to the fact that $SO(3)$ is not simply connected.

Given an initial point $x_0$ on a rotational fiber, other points $x$
on the same fiber can be parameterized by the rotation $R\in SO(3)$
such that $x=Rx_0$, or, equivalently, by the Euler angles of that
rotation.  Thus Euler angles become coordinates along a rotational
fiber.  We denote the Euler angles by $\theta^i$, $i=1,2,3$, or just
$\theta$ for short.  Shape or internal coordinates, on the other hand,
are rotationally invariant functions of $x$ or $\Xvec$.  We denote
these by $q^\mu$, $\mu=1,\ldots,3N-6$, or just $q$ for short.  These
can be taken to be coordinates along the section $S$, but, since they
are rotationally invariant, they are defined elsewhere in nuclear
configuration space by the fact that they are constant along
rotational fibers.

\subsection{Diabatic Bases}
\label{diabaticbases}

In multisurface problems the region of interest may include internal
degeneracies, usually conical intersections, and we may choose $x_0$
to lie on one of these in order to study frames in a corresponding
neighborhood.  Then small changes in shape as we move along $S$ away
from the conical intersection will break the degeneracy and produce an
eigenframe (an adiabatic basis) that is well defined but
discontinuous.  (The frame is smooth as $x\to x_0$ and the limit
exists, but the limit depends on the direction of approach.)  In such
cases the adiabatic basis has singularities regardless of phase or
frame conventions, and a smooth assignment of those conventions over a
section is impossible.  Therefore we must accept that an adiabatic
basis can be defined over a section $S$ and extended along rotational
fibers by (\ref{adiabphaseconv}), but that it will have
discontinuities as we vary the shape.  It will, however, be smooth as
we vary the orientation, holding the shape fixed.

The discontinuities in the adiabatic basis at degeneracies cause the
derivative couplings to diverge, and are the major drawback of this
basis.  To avoid these we may switch to a diabatic basis, an
orthonormal basis denoted $\ket{dx;k}$ with $d$ for ``diabatic.''  The
diabatic basis vectors for $k\in\Levels$ are required to span the subspace
$\Sspace(x)$, and to span $\Sspace^\perp(x)$ for $k\notin\Levels$.  In
addition, a diabatic basis is required to be smooth over its domain of
definition, something that can be achieved if we do not require the
basis vectors $\ket{dx;k}$ to be energy eigenstates for $k\in\Levels$.
Finally, we shall require diabatic bases to be invariant under time
reversal.

To construct a diabatic basis we begin with points $x_0$ on a section.
(In the following we use $x_0$ for a variable point in $S$.)  We
assume that a $T$-invariant, adiabatic basis $\ket{ax_0;k}$, including
phase and frame conventions, has been established for all $x_0$ in
some region in $S$.  As explained, this basis will have
discontinuities, in general.  Since sets of basis vectors,
$\ket{ax_0;k}$ and $\ket{dx_0;k}$ for $k\in\Levels$, are required to span
the same subspace $\Sspace(x)$, they must be related by
\begin{equation}
  \ket{dx_0;k} = \sum_{l\in\Levels} \ket{ax_0;l}\, V_{lk},
  \label{ADTdef}
\end{equation}
where $V_{kl}$ is an $N_l\times N_l$, unitary matrix that depends on
$x_0$.  Also, since $\ket{dx_0;k}$ is required to be smooth, the
matrix $V_{kl}$ must compensate for the singularities of the basis
$\ket{ax_0;k}$ and cannot be smooth itself.  Similar statements can
be made for the subset $k\notin\Levels$ and the complementary subspace
$\Sspace^\perp(x)$.

There are many ways to find the matrix $V$ so that the diabatic basis
is smooth.  Two of these, the singular-value basis and the
parallel-transported basis, have been discussed by us recently
(\cite{LittlejohnRawlinsonSubotnik22}).  The construction of both
bases takes place in a neighborhood of a fixed reference point
$x_{00}$ on $S$.  The construction is most interesting when $x_{00}$
lies on a degeneracy (a seam or conical intersection) but this is not
required.  The adiabatic and diabatic bases are required to agree at
$x_{00}$, $\ket{a,x_{00};k} = \ket{d,x_{00};k}$.  The singular value
diabatic basis is due to \cite{Pacheretal88, Pacheretal93}; it chooses
an orthonormal frame inside $\Sspace(x_0)$ for each $x_0$ in the
region of $S$ that is as close as possible to the adiabatic frame in
$\Sspace(x_{00})$, in the space of such frames.  The resulting field
of frames on $S$ is unique and smooth in a neighborhood of $x_{00}$
and defines the singular-value diabatic basis.  The
parallel-transported diabatic basis involves radial lines extending
out from $x_{00}$ to points $x_0\in S$, along which the basis is
carried by parallel transport.  This minimizes the distance in the
space of frames between the bases at $x_0$ and $x_0+dx_0$ for each
infinitesimal step along the curve.

Both the singular-value basis and the parallel-transported basis have
the property that if the adiabatic frame at $x_{00}$ is $T$-invariant,
as we assume, then so is the diabatic frame at all points $x_0$ in its
domain.  We omit the proofs but they involve the fact the projection
operator onto the strongly coupled subspace commutes with $T$, which
follows since that subspace is invariant under $T$. Thus, for these
diabatic bases, the matrix $V_{kl}$ in (\ref{ADTdef}) is real and
orthogonal.

Once the diabatic basis has been defined for $x_0\in S$, we extend the
definition along rotational fibers by means of rotation operators, as
in  (\ref{adiabphaseconv}).  Since we are assuming that the adiabatic
basis transforms by those same rotation operators, (\ref{ADTdef})
shows that the matrix $V_{kl}$ is independent of position along a
rotational fiber, that is, $V_{kl}(x)=V_{kl}(x_0)$, where $x=Rx_0$.
Equivalently, the matrix $V_{kl}$ is a function of shape coordinates
only and is independent of orientation.

\section{Transformation of Basis and Derivative Couplings Under Rotations}
\label{transfbasisF}

In this section we introduce the ``working basis,'' which is either
the adiabatic basis or a diabatic basis, whichever meets the
smoothness criteria.  It is the basis we will use for subsequent
calculations.  We continue with the electrostatic model.  We
accumulate a set of transformation laws for various quantities under
rotations, which supplement the transformation law (\ref{ESHeconjR})
for the electronic Hamiltonian and (\ref{epsilonkinv}) for its
eigenvalues, which we have already worked out.  The new transformation
laws include that for the basis vectors, (\ref{wbasisxfmrots}), that
for the matrix elements of the Hamiltonian, (\ref{Wklrotns}), and that
for the derivative couplings, (\ref{Fvecxfmrots}).  These are required
for subsequent work.

\subsection{The Working Basis and its Properties}
\label{workingbasis}

In the following we write simply $\ket{x;k}$ for a basis that is
either the adiabatic basis, in cases where that is smooth (for
example, in single-surface problems over a simply connected region),
otherwise a diabatic basis.  We will call this the ``working basis.''
We assume that it spans $\Sspace(x)$ for $k\in\Levels$ and
$\Sspace^\perp(x)$ for $k\notin\Levels$; that it is smooth over a section
$S$ or a chosen region thereof; that it is propagated along rotational
fibers by
\begin{equation}
  \ket{Rx_0;k}=U_{eo}(R)\,\ket{x_0;k};
  \label{basisrotxfm}
\end{equation}
and that it is $T$-invariant, $T\ket{x;k}=\ket{x;k}$.

Some authors, for example, \cite{Kendrick18}, have used a notation in
which the basis states are given as functions of the shape coordinates
alone, and not the Euler angles.  We believe it is worth clarifying
this, since it is obvious that the electronic eigenstates do depend on
orientation, and, with our phase conventions, (\ref{basisrotxfm}) shows
explicitly how they do.  On the other hand, points $x_0$ on the
section are determined by the shape coordinates, that is,
$x_0=x_0(q)$, so the basis states on the section can be regarded as
functions of shape coordinates alone.  We believe this is the correct
interpretation of something that in our notation would look like
$\ket{q;k}$, that is, it means $\ket{x_0(q);k}$.  There are objects
that really are constant along rotational fibers, for example, the
energy eigenvalues, which satisfy $\epsilon_k(x) =\epsilon_k(Rx_0)
=\epsilon_k(x_0) =\epsilon_k\bigl(x_0(q)\bigr) =\epsilon_k(q)$.  These
are functions of the shape coordinates alone everywhere in nuclear
configuration space, not just on the section.

We write the matrix elements of the electronic Hamiltonian in the
working basis for $k,l\in\Levels$ as
\begin{equation}
  \matrixelement{x;k}{H_e(x)}{x;l} = W_{kl}(x).
  \label{Wdef}
\end{equation}
If the working basis is the adiabatic basis then $W_{kl}(x) =
\epsilon_k(x)\,\delta_{kl}$, while in the diabatic basis $W_{kl}$ is a
full matrix.  In view of the $T$-invariance of the basis states
$\ket{x;k}$ the matrix $W_{kl}$ is real (hence, real and symmetric).
In view of the dependence (\ref{basisrotxfm}) of the basis states
along rotational fibers and the transformation law (\ref{ESHeconjR})
of the Hamiltonian, the matrix $W_{kl}$ is independent of orientation,
\begin{equation}
  W_{kl}(Rx_0) = W_{kl}(x_0),
  \label{Wklrotns}
\end{equation}
that is, $W_{kl}$ depends only on the shape coordinates.

We turn now to the transformation properties of the working basis
under rotations.  Let $x_0\in S$ be an initial point on a rotational
fiber, let $R_1,R_2 \in SO(3)$, and let $x_1=R_1x_0$ and $x_2=R_2x_1$.
Then
\begin{eqnarray}
  \ket{x_2;k} &=& \ket{R_2x_1;k} =\ket{R_2R_1x_0;k}
  =  U_{eo}(R_2R_1)\,\ket{x_0;k}\nonumber\\
  &=& U_{eo}(R_2)\,U_{eo}(R_1)\,\ket{x_0;k}
  =U_{eo}(R_2)\,\ket{x_1;k},
\end{eqnarray}
where we use (\ref{basisrotxfm}) and the representation property
(\ref{Ueorepn}).  Making the notational changes, $x_1\to x$, $R_2\to
R$, we can write this as
\begin{equation}
  \ket{Rx;k} = U_{eo}(R)\,\ket{x;k}.
  \label{wbasisxfmrots}
\end{equation}

This is the transformation law of the working basis $\ket{x;k}$ under
rotations.  It is the same as (\ref{basisrotxfm}) with $x_0$ replaced
by $x$, but the meaning is quite different.  That is,
(\ref{basisrotxfm}) defines the working basis along the rotational
fibers, including the phase and frame conventions, given such a
definition at an initial point $x_0\in S$; whereas
(\ref{wbasisxfmrots}) is a property of those basis vectors, once
defined, at any point $x$ on a rotational fiber.  (Of course,
(\ref{basisrotxfm}) is a special case of (\ref{wbasisxfmrots}).)

Let us now specialize $R$ in (\ref{wbasisxfmrots}) to an infinitesimal
rotation, that is, one for which the angle $\theta$ is infinitesimal,
so that
\begin{equation}
  R=R(\nvechat,\theta)= I+\theta\nvechat\times.
  \label{infinitesR}
\end{equation}
If we let such an $R$ act on a configuration $x$, then we can write
$Rx=x+\delta x$, where
\begin{equation}
  \delta x =\theta(\nvechat\times\Xvec_1,\ldots,
  \nvechat\times\Xvec_{N-1}).
  \label{deltaxdef}
\end{equation}
On the other hand, when $\theta$ is small, (\ref{Ueodef}) implies
\begin{equation}
  U_{eo}(R) = 1-\frac{i}{\hbar}\,\theta\nvechat\cdot\Lvec_e.
  \label{infinitesUeo}
\end{equation}
For such rotations (\ref{wbasisxfmrots}) becomes
\begin{equation}
  \ket{Rx;k} = \ket{x+\delta x;k} =\ket{x;k}
  + \sum_{\alpha=1}^{N-1} \theta(\nvechat\times\Xvec_\alpha)\cdot
  \nabla_\alpha\ket{x;k} 
  =\ket{x;k}-\frac{i}{\hbar}\,\theta\nvechat\cdot\Lvec_e\,\ket{x;k}.
  \label{deltaRbasis}
\end{equation}
In this we write $(\nvechat\times\Xvec_\alpha)\cdot\nabla_\alpha =
\nvechat\cdot(\Xvec_\alpha\times\nabla_\alpha)$, we cancel leading
terms, the factor $\theta$ and the factor $\nvechat$ (which is an
arbitrary unit vector).  The result can be written as
\begin{equation}
  \left[-i\hbar\sum_{\alpha=1}^{N-1} (\Xvec_\alpha \times
    \nabla_\alpha)+ \Lvec_e\right]\ket{x;k} =0,
\end{equation}
or,
\begin{equation}
  (\Lvec_n + \Lvec_e)\ket{x;k}=0,
  \label{Ltotbasiszero}
\end{equation}
where $\Lvec_n$ is the usual differential operator for the nuclear
orbital angular momentum, here acting on the parametric dependence $x$
of the working basis states. 

In the case of nondegenerate adiabatic basis states, \cite{Yarkony01}
has shown that the off-diagonal matrix elements of $\Lvec_n+\Lvec_e$
vanish (see Eq.~(19abc) of that article).  Because of the
time-reversal invariance of the basis states, the diagonal elements
vanish, too, and Yarkony's results are equivalent to
(\ref{Ltotbasiszero}).  With our phase and frame conventions, however,
(\ref{Ltotbasiszero}) applies also in the case of degeneracies or
diabatic bases.

\subsection{Transformation Properties of Derivative Couplings}
\label{derivcouprotn}

We wish to find how the derivative couplings, defined by (\ref{Fvecdef}),
transform along rotational fibers, that is, how
$\Fvec_{\alpha;kl}(Rx)$ depends on $\Fvec_{\alpha;kl}(x)$.  The
obvious strategy is to work with $\Fvec_{\alpha;kl}(Rx)
=\matrixelement{Rx;k}{\nabla_\alpha}{Rx;l}$, but this is notationally
awkward (do we rotate first and then differentiate, or the other way
around?)

Therefore we propose a different approach.  Let $\xi$ be an
infinitesimal displacement in nuclear configuration space,
\begin{equation}
  \xi = (\xivec_1,\ldots,\xivec_{N-1}).
  \label{xidef}
\end{equation}
Then we have
\begin{equation}
  \ket{x+\xi;l}=\ket{x;l}+\sum_{\alpha=1}^{N-1}\xivec_\alpha
  \cdot\nabla_\alpha\ket{x;l},
\end{equation}
which implies
\begin{equation}
  \braket{x;k}{x+\xi;l} = \delta_{kl} +
  \sum_{\alpha=1}^{N-1} \xivec_\alpha\cdot\Fvec_{\alpha;kl}(x).
\end{equation}
The $x$ in this equation is a dummy variable, and one that is
independent of $\xi$, so we can replace it by $Rx$, obtaining
\begin{equation}
  \braket{Rx;k}{Rx+\xi;l} = \delta_{kl} +
  \sum_{\alpha=1}^{N-1} \xivec_\alpha\cdot\Fvec_{\alpha;kl}(Rx).
  \label{FvecxtoRx}
\end{equation}
On the other hand, we have
\begin{eqnarray}
  \ket{Rx+\xi;l} &=& \ket{R(x+R^{-1}\xi);l}
    =U_{eo}(R)\,\ket{x+R^{-1}\xi;l}\nonumber\\
    &=& U_{eo}(R)\left[\ket{x;l}
      +\sum_{\alpha=1}^{N-1}(R^{-1}\xivec_\alpha)\cdot
      \nabla_\alpha\ket{x;l}\right],
    \label{Rxxiright}
\end{eqnarray}
and
\begin{equation}
  \bra{Rx;k} = \bra{x;k}\,U_{eo}(R)^\dagger,
  \label{Rxxileft}
\end{equation}
where we use (\ref{wbasisxfmrots}).  Upon taking the product of
(\ref{Rxxileft}) and (\ref{Rxxiright}) the rotation operators cancel
and we obtain
\begin{equation}
  \braket{Rx;k}{Rx+\xi;l}=\delta_{kl} + \sum_{\alpha=1}^{N-1}
  \xivec_\alpha \cdot (R\Fvec_{\alpha;kl}(x)),
\end{equation}
where we have used the fact that for any two vectors $\Avec$ and
$\Bvec$ and any rotation $R$, we have $(R^{-1}\Avec)\cdot\Bvec =
\Avec\cdot (R\Bvec)$.  Comparing this with (\ref{FvecxtoRx}), we
obtain finally
\begin{equation}
  \Fvec_{\alpha;kl}(Rx) = R\Fvec_{\alpha;kl}(x),
  \label{Fvecxfmrots}
\end{equation}
since the displacements $\xivec_\alpha$ are arbitrary.

Equation~(\ref{Fvecxfmrots}) is the transformation law of the
derivative couplings under rotations; it says, in a sense, that
$\Fvec_{\alpha;kl}$ transforms as a vector field on nuclear
configuration space under rotations.  This result is important for
establishing the rotational invariance of the Born-Oppenheimer
Hamiltonian for multisurface, electrostatic problems.

Some authors, for example, \cite{Yarkony01}, have used a notation in
which the derivative couplings are given as functions of the shape
coordinates alone, with no dependence on the Euler angles.  It would
be as if we wrote $\Fvec_{\alpha;kl}(q)$ in our notation.  We believe
this means $\Fvec_{\alpha;kl}\bigl(x_0(q)\bigr)$, that is, it is the
derivative couplings evaluated on the section, where $q^\mu$ are
coordinates.  The derivative couplings elsewhere do depend on the
orientation, as shown explicitly by (\ref{Fvecxfmrots}).

A subtlety in this matter is that the derivative couplings are really
the components of a differential form with respect to the Jacobi
vectors, and if the components are taken instead with respect to
rotationally invariant vector fields, then those components will be
rotationally invariant.  This is an issue in the construction of
kinetic energy operators in the internal space that incorporate
geometric phase effects.

\subsection{Rotational Components of Derivative Couplings}
\label{rotcompsFvec}

Let $R$ be an infinitesimal rotation with $\theta\ll 1$, and let
$Rx=x+\delta x$, as in (\ref{infinitesR}) and (\ref{deltaxdef}).  Then
the component of the derivative couplings in the direction $\delta x$
is
\begin{eqnarray}
  \sum_{\alpha=1}^{N-1} \delta\Xvec_\alpha\cdot\Fvec_{\alpha;kl}(x)
  &=&\theta \sum_{\alpha=1}^{N-1} 
  \matrixelement{x;k}{(\nvechat\times\Xvec_\alpha)\cdot\nabla_\alpha}
  {x;l}
  =\frac{i}{\hbar}\,\theta\nvechat\cdot
  \sum_{\alpha=1}^{N-1} \matrixelement{x;k}{\Xvec_\alpha\times
    \Pvec_\alpha}{x;l}\nonumber\\
  &=&-\frac{i}{\hbar}\,\theta\nvechat\cdot
  \matrixelement{x;k}{\Lvec_e}{x;l},
  \label{Frotnlcomp}
\end{eqnarray}
where in the last step we use (\ref{Ltotbasiszero}).  Thus, the
angular components of the derivative couplings (with our phase and
frame conventions) are proportional to the matrix elements of the
electronic angular momentum.  This result is due to \cite{Yarkony01};
it is important for the construction of kinetic energy operators on the
internal space.

\section{Molecular and Born-Oppenheimer Representations}
\label{molecBOreps}

We now introduce the molecular and Born-Oppenheimer representations for
the state space of the molecule, which give us the setting within
which our main results concerning angular momentum can be stated.   We
continue with the electrostatic model, with possibly more than one
strongly coupled potential energy surface.

\subsection{Two Representations}
\label{tworeps}

In what we call the ``molecular representation,'' the quantum state of
the molecule is specified by the wave function $\Psi(\Xvec,\rvec)$,
which is just standard quantum mechanics on the standard Hilbert space
for the molecule.  If $\Psi$ is any such wave function, we expand the
$\rvec$ dependence in terms of the $x$-dependent working basis vectors
$\ket{x;k}$, whose wave functions are denoted $\phi_k(\Xvec;\rvec)$,
as shown in (\ref{phiket}).  That is, we write
\begin{equation}
  \Psi(\Xvec,\rvec)=\sum_k \psi_k(\Xvec)\,\phi_k(\Xvec;\rvec),
  \label{Psiintermsofpsi}
\end{equation}
as is standard in Born-Oppenheimer theory.  Here $\psi_k(\Xvec)$ are
the expansion coefficients; we imagine them forming an
infinite-dimensional vector of wave functions of $\Xvec$.  Equation
(\ref{Psiintermsofpsi}) gives the wave function $\Psi(\Xvec,\rvec)$ in
terms of the purely nuclear wave functions $\psi_k(\Xvec)$; the
inverse relation is
\begin{equation}
  \psi_k(\Xvec) = \int d\rvec\,\phi_k(\Xvec;\rvec)^*
  \, \Psi(\Xvec,\rvec).
  \label{psiintermsofPsi}
\end{equation}
We will refer to the infinite dimensional vector of nuclear wave
functions $\psi_k(\Xvec)$ as the ``Born-Oppenheimer representation''
of the quantum state of the molecule, and abbreviate the relationships
(\ref{Psiintermsofpsi}) and (\ref{psiintermsofPsi}) between them by
writing
\begin{equation}
  \Psi(\Xvec,\rvec) \longleftrightarrow \psi_k(\Xvec).
  \label{Psiarrowpsi}
\end{equation}
The association is one-to-one, and no information is lost by using the
Born-Oppenheimer representation.

Similarly, let $A$ be a linear operator that maps molecular wave
functions $\Psi(\Xvec,\rvec)$ into new such wave functions
$\Psi'(\Xvec,\rvec)$, something we can write as $\Psi'(\Xvec,\rvec) =
(A\Psi)(\Xvec,\rvec)$.   This is in the molecular representation.   In
the Born-Oppenheimer representation, $A$ is replaced by an
infinite-dimensional matrix $A_{kl}$ of linear operators, each of
which acts on wave functions $\psi(\Xvec)$, depending on the nuclear
coordinates alone.  That is, if $\Psi'=A\Psi$ as shown, and if
$\Psi\longleftrightarrow \psi_k$ and $\Psi'\longleftrightarrow
\psi'_k$, then
\begin{equation}
  \psi'_k(\Xvec) = \sum_l (A_{kl}\,\psi_l)(\Xvec).
  \label{psi'Aklpsi}
\end{equation}
This is equivalent to
\begin{equation}
  (A\Psi)(\Xvec,\rvec) = \sum_k \phi_k(\Xvec;\rvec)
  \sum_l (A_{kl}\,\psi_l)(\Xvec),
  \label{APsiAklpsil}
\end{equation}
which gives $A$ in terms of the matrix of operators $A_{kl}$.  The
inverse is
\begin{equation}
  (A_{kl}\psi)(\Xvec) = \int d\rvec \, \phi_k(\Xvec;\rvec)^*
  \, (A\Phi_l)(\Xvec,\rvec),
  \label{AklintermsofA}
\end{equation}
where $\Phi_l(\Xvec,\rvec) = \psi(\Xvec)\,\phi_l(\Xvec;\rvec)$.  We
write $\psi$ without a subscript in (\ref{AklintermsofA}) because it
is just a dummy function of $\Xvec$ that is used to define the
operator $A_{kl}$; it may be the component of a wave function in the
Born-Oppenheimer representation, but it need not be.  We will summarize
the relations (\ref{APsiAklpsil}) and (\ref{AklintermsofA}) between
the operators in the two representations by writing
\begin{equation}
  A \longleftrightarrow A_{kl}.
  \label{AarrowAkl}
\end{equation}

Then it is easy to prove some theorems.  If $B=A^\dagger$ (in the
molecular representation), then in the Born-Oppenheimer representation
we have
\begin{equation}
  B_{kl} = (A_{lk})^\dagger,
  \label{ABdaggerrule}
\end{equation}
where the parentheses make it clear that we form the transpose of the
matrix $A_{kl}$ first, and then take the Hermitian conjugate of the
elements.

Likewise, if $A$, $B$ and $C$ are operators in the molecular
representation and $C=AB$, then
\begin{equation}
  C_{kl} = \sum_p A_{kp}\,B_{pl},
  \label{CABrule}
\end{equation}
that is, operator products are mapped into matrix products (but the
matrix elements are themselves operators, and their order of
multiplication must be respected).

\subsection{Examples of Operators in the Born-Oppenheimer Representation}
\label{examplesops}

We present some examples of the transformation of operators from the
molecular representation to the Born-Oppenheimer representation.  If
$A=f(x)$ is a function of $x$ in the molecular representation, that
is, a multiplicative operator on wave functions $\Psi(x,\rvec)$,  then
we find
\begin{equation}
  f(x) \longleftrightarrow f(x)\,\delta_{kl}.
  \label{f(x)BO}
\end{equation}
In particular, this applies when $f$ is one of the components of the
Jacobi vectors $\Xvec_\alpha$.  

In the following we let $\Pvec_\alpha$ stand for the differential
operator $-i\hbar\nabla_\alpha$.   In the molecular representation,
this represents physically the kinetic momentum conjugate to the
Jacobi vector $\Xvec_\alpha$.  Transforming to the Born-Oppenheimer
representation, we find
\begin{equation}
  \Pvec_\alpha \longleftrightarrow \Pvec_\alpha \, \delta_{kl}
  -i\hbar\,\Fvec_{\alpha;kl}(x).
  \label{PalphaBO}
\end{equation}
The Born-Oppenheimer version of this operator has a well known
interpretation as a covariant derivative (\cite{Bohmetal91}).   

For another example, consider a purely electronic operator, for
example, $\Lvec_e$.   Then we find
\begin{equation}
  \Lvec_e \longleftrightarrow \matrixelement{x;k}{\Lvec_e}{x;l},
  \label{LeBO}
\end{equation}
that is, the matrix of nuclear operators representing $\Lvec_e$ in the
Born-Oppenheimer representation are purely multiplicative functions of
$x$, which are otherwise the matrix elements of $\Lvec_e$ in the
working basis. Another purely electronic operator is the electronic
Hamiltonian, which, however, depends on $x$.  We have
\begin{equation}
  H_e(x) \longleftrightarrow W_{kl}(x),
  \label{HeBO}
\end{equation}
see (\ref{Wdef}) and (\ref{Wklrotns}).

Yet another such operator is the projector onto the strongly coupled
subspace $\Sspace(x)$,
\begin{equation}
  P(x) = \sum_{k\in\Levels} \ketbra{x;k}{x;k}.
  \label{P(x)def}
\end{equation}
This maps into its Born-Oppenheimer version, $P(x) \longleftrightarrow
P_{kl}$, where
\begin{equation}
  P_{kl} = \begin{cases}
    \delta_{kl}, & \text{if $k,l\in\Levels$}\\
    0, & \text{otherwise.}
    \end{cases}
  \label{Pkldef}
\end{equation}

\subsection{The Molecular Hamiltonian}
\label{HmolBOrepn}

The molecular Hamiltonian in the electrostatic model and in the
molecular representation is
\begin{equation}
  H_{\rm mol} = \sum_{\alpha=1}^{N-1}\frac{\Pvec_\alpha^2}{2M_\alpha}
  + H_e(x).
  \label{HmolESdef}
\end{equation}
Transforming this to the Born-Oppenheimer representation, $H_{\rm mol}
\longleftrightarrow H_{{\rm mol},kl}$, we find
\begin{equation}
  H_{{\rm mol},kl} = \sum_{\alpha=1}^{N-1}\left\{
  \frac{1}{2M_\alpha}\sum_p 
  [\Pvec_\alpha\,\delta_{kp} -i\hbar\,\Fvec_{\alpha;kp}(x)]\cdot
  [\Pvec_\alpha\,\delta_{pl} -i\hbar\,\Fvec_{\alpha;pl}(x)]
  \right\}+W_{kl}(x),
    \label{Hmolkldef}
\end{equation}
where we use the product rule (\ref{CABrule}), (\ref{PalphaBO}) and
(\ref{HeBO}).  This style of transforming to the Born-Oppenheimer
representation follows \cite{Kendrick18}.  The sum
on $p$ is the matrix multiplication indicated by (\ref{CABrule});
notice that this sum runs over the entire Hilbert space, that is,
both $p\in\Levels$ and $p\notin\Levels$.  

In most physical circumstances of interest the nuclear momentum
$\Pvec_\alpha$ is large when measured in atomic units, because of the
large nuclear mass, while $\Fvec_{\alpha;kl}$ (with our phase
conventions and smoothness assumptions) is of order unity in the same
units.  Therefore the terms of the kinetic energy in (\ref{Hmolkldef})
decrease in magnitude as the power of $\Pvec_\alpha$ decreases.
Therefore the diagonal terms $k=l$ are dominated by $\Pvec_\alpha^2$
while the off-diagonal terms $k\ne l$ are dominated by the terms
linear in $\Pvec_\alpha$, which are therefore smaller than the
diagonal terms.

We can decouple the strongly coupled levels $k\in\Levels$ from the
rest by simply throwing away the off-diagonal terms $(kl)$ of
(\ref{Hmolkldef}) for $k\in\Levels$ and $l\notin\Levels$ or
$k\notin\Levels$ and $l\in\Levels$.  This replaces the Hamiltonian
$H_{{\rm mol},kl}$ by a new, block-diagonal, one that we will call
$K_{{\rm mol},kl}$.  The formula for the latter is the same as
(\ref{Hmolkldef}) when $k,l\in\Levels$ or $k,l\notin\Levels$, and 0
otherwise.  Alternatively, since we do not care about dynamics outside
the strongly coupled subspace, we can define $K_{{\rm mol},kl}$ as the
same as $H_{{\rm mol},kl}$ when $k,l\in\Levels$, and 0 otherwise.
This is equivalent to
\begin{equation}
  K_{\rm mol} = P(x) H_{\rm mol} P(x),
  \label{HbarP(x)def}
\end{equation}
that is, it is just the original molecular Hamiltonian, projected onto
the strongly coupled subspace.  Most derivations of the
Born-Oppenheimer Hamiltonian in the literature amount to carrying out
this projection.

In the special case of a single surface problem, where $\Levels$
contains the single level $k_0$, this procedure gives us the
Born-Oppenheimer Hamiltonian (\ref{HBOdef}) with $k\to k_0$.  The
latter would be written in the present notation as $K_{{\rm
    mol},k_0k_0}$.  As noted, the derivative couplings
$\Fvec_{\alpha;kl}$ vanish on the diagonal $(kl)=(k_0k_0)$.  In this
context the replacement of $H_{\rm mol}$ by $K_{\rm mol}$ is
usually called ``the Born-Oppenheimer approximation.''  Its obvious
generalization to multisurface problems is given by
(\ref{HbarP(x)def}).

Actually, the ``Born-Oppenheimer approximation'' is often described
(in the context of a single-surface problem) as one in which the wave
function is assumed to have the product form seen in
(\ref{Psipsiphi}).  But if the Hamiltonian is approximated by throwing
away off-diagonal terms, then the new Hamiltonian possesses solutions
of the product form.  Therefore we regard the usual Born-Oppenheimer
approximation as one of approximating the Hamiltonian.

The off-diagonal terms that we throw away to obtain $K_{{\rm mol},kl}$
are indeed small compared to the diagonal terms, but beyond this their
neglect is {\em ad hoc} and it is hard to find a deeper justification
for the procedure in the literature, at least in the case of
large-amplitude motions.  This leaves open the question of whether the
Born-Oppenheimer Hamiltonians obtained by projection as in
(\ref{HbarP(x)def}) are even correct.  They certainly are so to first
order in small quantities, but it is not obvious that they are correct
to second order.  In Sec.~\ref{dressingxfm} we will discuss Moyal
perturbation theory, which is useful for answering these questions.
The issue is more important than the small, second order terms in the
Hamiltonian, as it involves the dressing of the nuclear variables,
which has effects at first order.

\subsection{The Angular Momentum}
\label{angmomborepn}

In the molecular representation the total orbital angular momentum of
the molecule is represented by the operator $\Lvec=\Lvec_n+\Lvec_e$,
which is given by (\ref{Lvecdef}).  To find the Born-Oppenheimer
representation we allow $\Lvec$ to act on a molecular wave function,
\begin{eqnarray}
  \Lvec\Psi(\Xvec,\rvec) &=& (\Lvec_n + \Lvec_e)
  \sum_k \psi_k(\Xvec)\,\phi_k(\Xvec;\rvec)\nonumber\\
  &=& \sum_k[(\Lvec_n+\Lvec_e)\psi_k(\Xvec)]\,\phi_k(\Xvec;\rvec)+
  \sum_k\psi_k(\Xvec)\,[(\Lvec_n+\Lvec_e)\phi_k(\Xvec;\rvec)],
  \label{Ldistribute}
\end{eqnarray}
where we distribute $\Lvec_n+\Lvec_e$ using the product or Leibnitz
rule, since it is a first-order, linear, differential operator.  Then
the second major sum on the right vanishes due to
(\ref{Ltotbasiszero}), while in the first sum on the right the term
involving $\Lvec_e$ also vanishes, since $\psi_k$ has no dependence on
$\rvec$.   The result can be written,
\begin{equation}
  \Lvec=\Lvec_n+\Lvec_e \longleftrightarrow \Lvec_n\,\delta_{kl}.
  \label{LvecBOrepn}
\end{equation}
This justifies and makes precise our earlier statement that in the
Born-Oppenheimer representation, the nominal, nuclear orbital angular
momentum includes physically both the nuclear and electronic orbital
angular momenta, and that this is exact.

The molecular Hamiltonian $H_{\rm mol}$ in the molecular
representation commutes with $\Lvec$, due to the overall rotational
invariance of the molecule, so in the Born-Oppenheimer representation
the matrix $H_{{\rm mol},kl}$ must commute with the matrix
$\Lvec_n\,\delta_{kl}$.  But since the latter is a multiple of
the identity, this reduces to
\begin{equation}
  [\Lvec_n,H_{{\rm mol},kl}]=0.
  \label{HmolLncomrel}
\end{equation}
This is the form that overall angular momentum conservation takes in
the Born-Oppenheimer representation; every component of the matrix
$H_{{\rm mol},kl}$ commutes with $\Lvec_n$, that is, it is a scalar
under nuclear orbital rotations.

Nuclear orbital rotations are generated by $\Lvec_n$ and are
implemented by the operators,
\begin{equation}
  U_{no}(R)=U_{no}(\nvechat,\theta) = \exp\left(-\frac{i}{\hbar}\theta
    \nvechat\cdot\Lvec_n\right),
  \label{Unodef}
\end{equation}
which act on nuclear wave functions according to $\bigl(
U_{no}(R)\psi\bigr) (\Xvec) =\psi\bigl (R^{-1}\Xvec\bigr)$.  Like the
electronic orbital rotation operators $U_{eo}(R)$, nuclear orbital
rotation operators form a representation of $SO(3)$,
$U_{no}(R_1)\,U_{no}(R_2) = U_{no}(R_1R_2)$.  An operator commutes
with nuclear orbital angular momentum $\Lvec_n$ if and only if it
commutes with the rotations $U_{no}(R)$ for all $R\in SO(3)$.
Therefore to check (\ref{HmolLncomrel}) we can see how $H_{{\rm
    mol},kl}$ transforms under conjugation by $U_{no}(R)$.

To begin we have
\begin{eqnarray}
	U_{no}(R)\,\Xvec_\alpha \,U_{no}(R)^\dagger &=&
	R^{-1}\Xvec_\alpha,\label{Xalphaconj}\\
	U_{no}(R)\,\Pvec_\alpha \,U_{no}(R)^\dagger &=&
	R^{-1}\Pvec_\alpha,\label{Palphaconj}
\end{eqnarray}
which is a statement that $\Xvec_\alpha$ and $\Pvec_\alpha$ are vector
operators.  This implies that
$\Pvec_\alpha^2=\Pvec_\alpha\cdot\Pvec_\alpha$ is a scalar.  Next, we
have
\begin{equation}
	U_{no}(R) \, \Fvec_{\alpha;kl}(x)\,U_{no}(R)^\dagger
	= \Fvec_{\alpha;kl}\bigl(R^{-1}x\bigr)
	=R^{-1}\Fvec_{\alpha;kl}(x),
	\label{Fvecconjrel}
\end{equation}
where in the first step we use (\ref{Xalphaconj}) and in the second,
(\ref{Fvecxfmrots}).  This shows that the derivative couplings, with
our choice of basis states and phase conventions, transform under
nuclear rotations as a vector operator.  Thus, dot products such as
$\Pvec_\alpha\cdot\Fvec_{\alpha;kl}$ are scalars.  We see that the
components $H_{{\rm mol},kl}$ are scalars, and that therefore they
commute with $\Lvec_n$.  The same is true for the components $K_{{\rm
mol},kl}$, which are either equal to $H_{{\rm mol},kl}$ or else are
zero.

In single-surface problems it is obvious that the Born-Oppenheimer
version of the Hamiltonian, (\ref{HBOdef}), commutes with $\Lvec_n$,
because it has the simple kinetic-plus-potential form with a
rotationally invariant potential and there are no derivative
couplings.  This means that energy eigenfunctions can be organized as
eigenfunctions also of $L_n^2$ and $L_{nz}$, as noted.

In multisurface problems, as we have just shown, the matrix
Hamiltonian $K_{{\rm mol},kl}$ commutes with the matrix of angular
momentum operators, $\Lvec_n\,\delta_{kl}$. This means that the
solutions of the Born-Oppenheimer version of the Schr\"odinger
equation, which now reads (for $k\in\Levels$),
\begin{equation}
	\sum_{l\in\Levels} K_{{\rm mol},kl}\,\psi_l(\Xvec)
	=E\,\psi_k(\Xvec),
	\label{BOScheqnmultilevel}
\end{equation}
can be organized as simultaneous eigenfunctions of the matrix
Hamiltonian $K_{{\rm mol},kl}$ and  the matrix angular momentum
operators, $L_n^2\,\delta_{kl}$ and $L_{nz}\, \delta_{kl}$.  But
to make the vector wave function $\psi_k$ for $k\in\Levels$ an
eigenfunction of those  matrix angular momentum operators, each
component $\psi_k(\Xvec)$ must be an eigenfunction of $L_n^2$ and
$L_{nz}$ with the same quantum numbers.  Call these $(l,m_l)$.

We then transform the eigenfunction $\psi_k(\Xvec)$, which is nonzero
only for $k\in\Levels$, back to the molecular representation, using
(\ref{Psiintermsofpsi}) but only summing over $k\in\Levels$.  The
resulting molecular wave function $\Psi(\Xvec,\rvec)$ is then an exact
eigenfunction of $\Lvec^2$ and $L_z$, where $\Lvec=\Lvec_n+\Lvec_e$,
with the same quantum numbers $(l,m_l)$.  Thus we obtain an
understanding of angular momentum conservation in multi-surface
problems. 

\section{Details in Fine Structure Models}
\label{detailsFS}

Some of the changes required on passing from the electrostatic model
to the fine structure model have been discussed at the beginning of
Sec.~\ref{answersfs}.  These include the facts that the electronic and
molecular wave functions, $\phi(\rvec, m)$ and $\Psi(\Xvec,\rvec,m)$,
respectively, acquire a dependence on the spin quantum numbers $m$
(see (\ref{mdef})); that the electronic and molecular Hamiltonians,
$H_e(x;\rvec,\pvec,\Svec)$ and $H_{\rm mol} (\Xvec ,\Pvec ,\rvec
,\pvec,\Svec)$, respectively, acquire a dependence on the electron
spin $\Svec$; and that the definition of time reversal changes, from
(\ref{TdefES}) to (\ref{TdefFSabbrev}).

The electronic Hamiltonian $H_e(x;\rvec,\pvec,\Svec)$ now depends on
the dot products and triple products of the vectors $\Xvec_\alpha$,
$\rvec_i$, $\pvec_i$ and $\Svec_i$, so (\ref{ESH_einvR}) is replaced
by
\begin{equation}
  H_e(x;\rvec,\pvec,\Svec) = H_e(Rx;R\rvec,R\pvec,R\Svec)
  \qquad \forall R\in SO(3),
  \label{HFS_einvR}
\end{equation}
where $R\Svec=(R\Svec_1,\ldots,R\Svec_{N_e})$.  This is a statement
about the functional form of the electronic Hamiltonian.  

To connect this with rotation operators we cannot use orbital
rotations as in (\ref{ESHeconjR}) but rather we must introduce total
electron rotation operators that include the spin.  We denote these by
$U_e(u)$; they are parameterized by an element $u\in SU(2)$ or by the
equivalent axis and angle, $U_e(\nvechat,\theta)
=U_e\bigl(u(\nvechat,\theta)\bigr)$, and are defined by their action
on electronic wave functions,
\begin{equation}
  \bigl(U_e(u)\phi\bigr)(\rvec,m) =
  \sum_{m'} (u\ldots u)_{mm'}\,\phi\bigl(R^{-1}\rvec,m'\bigr),
  \label{Ueaction}
\end{equation}
where the notation for the sum is the same as in (\ref{TdefFSabbrev})
and where $R$ means $R(u)$, defined by (\ref{R(u)def}).  It follows
from (\ref{Ueaction}) that the operators $U_e(u)$ form a
representation of $SU(2)$,
\begin{equation}
  U_e(u_1)U_e(u_2) = U_e(u_1u_2).
  \label{UeSU2repn}
\end{equation}
These operators are given in terms of their generators by
\begin{equation}
  U_e(u) = U_e(\nvechat,\theta) = \exp\left[-\frac{i}{\hbar}
      \,\theta\nvechat\cdot(\Lvec_e+\Svec)\right].
  \label{Uedef}
\end{equation}

Now all of $\rvec$, $\pvec$ and $\Svec$ transform as vector operators
under conjugation by $U_e(u)$, for example, we have
\begin{equation}
  U_e(u)\,\Svec_i \,U_e(u)^\dagger = R^{-1}\Svec_i,
  \label{SvecUeconj}
\end{equation}
where in formulas like this it is understood that $R=R(u)$.  Therefore
the electronic Hamiltonian transforms according to
\begin{equation}
  U_e(u) \, H_e(x;\rvec,\pvec,\Svec)\,U_e(u)^\dagger
  = H_e\bigl(x;R^{-1}\rvec,R^{-1}\pvec,R^{-1}\Svec\bigr)
  = H_e(Rx;\rvec,\pvec,\Svec),
\end{equation}
just as in the derivation of (\ref{ESHeconjR0}).  With the
abbreviation $H_e(x;\rvec,\pvec,\Svec)\to H_e(x)$ this becomes
\begin{equation}
  U_e(u) \, H_e(x) \, U_e(u)^\dagger = H_e(Rx),
  \label{FSHeconjR}
\end{equation}
which may be compared to its electrostatic counterpart,
(\ref{ESHeconjR}).  They are the same except that $U_{eo}(R)$ has been
replaced by $U_e(u)$.  This is the transformation law of the
electronic Hamiltonian along rotational fibers in the fine structure
model.

\subsection{Fine Structure Details, Even Number of Electrons}
\label{FSdetailseven}

Most of the results in the fine structure model with $N_e={\rm even}$
have the same form as in the electrostatic model, since the main
conclusions follow from $T^2=+1$ which holds in both cases (although
the definitions of $T$ are not the same).  We begin with rotation
operators.  

The function $R(u)$ (see (\ref{R(u)def})) has the property
$R(u)=R(-u)$ and if the number of electrons is even then the number of
factors of $u$ in (\ref{Ueaction}) is also even.  Thus
$U_e(u)=U_e(-u)$ and $U_e$ might as well be parameterized by
$R=R(u)\in SO(3)$.  Then the operators $U_e(R)$ form a representation
of $SO(3)$, $U_e(R_1)U_e(R_2)=U_e(R_1R_2)$.  The definition of
$U_e(R)$ is still (\ref{Ueaction}), but with $U_e(R)$ instead of
$U_e(u)$ on the left hand side and with the understanding that $u$ on
the right hand side is one of the two elements of $SU(2)$ that
correspond to the given $R\in SO(3)$ according to (\ref{R(u)def}).
These differ by a sign, which the answer does not depend on.

Thus, the transformation law for the Hamiltonian along rotational
fibers in the fine structure model with an even number of electrons is
$U_e(R)\,H_e(x)\,U_e(R)^\dagger = H_e(Rx)$, which may be compared to
(\ref{ESHeconjR}) in the electrostatic model (they are the same,
except the rotation operator now includes spin).

As for the basis states, we still have $T^2=+1$ as in the
electrostatic model and $T$ still commutes with $H_e(x)$, which
together imply the existence of $T$-invariant bases (adiabatic first,
and then diabatic).  These can be defined along a section $S$,
smoothly, in the case of the diabatic basis, and then propagated along
noncollinear rotational fibers by
\begin{equation}
  \ket{Rx_0;k} = U_e(R)\,\ket{x_0;k}.
  \label{basisrotxfmFSeven}
\end{equation}
This may be compared to its electrostatic counterpart,
(\ref{basisrotxfm}); the only difference is that the rotation operator
now includes spin.

With these (smooth) phase and frame conventions we can define a smooth
working basis, as in the electrostatic model.   This basis transforms
under rotations according to
\begin{equation}
	\ket{Rx;k}=U_e(R)\,\ket{x;k},
	\label{wbasisxfmrotsFSeven}
\end{equation}
which is just like (\ref{wbasisxfmrots}) and proved in the same way,
except that the rotation now involves spin.  By making $R$
infinitesimal in this, we obtain
\begin{equation}
	(\Lvec_n+\Lvec_e+\Svec)\ket{x;k}=0,
	\label{JtotbasiszeroFSeven}
\end{equation}
just like (\ref{Ltotbasiszero}) except that now the spin is included.
The operator that appears is the total angular momentum $\Jvec$ of the
molecule.  

We then find that the derivative coupling transform under rotations
according to (\ref{Fvecxfmrots}), the same formula as in the
electrostatic model.  As for the rotational components of the
derivative couplings, they are now given by
\begin{equation}
  \sum_{\alpha=1}^{N-1} \delta\Xvec_\alpha\cdot\Fvec_{\alpha;kl}(x)
  = -\frac{i}{\hbar}\,\theta\nvechat\cdot
  \matrixelement{x;k}{(\Lvec_e+\Svec)}{x;l},
  \label{FrotnlcompFSeven}
\end{equation}
where $\delta\Xvec_\alpha=\theta\nvechat\times\Xvec_\alpha$ and where
the matrix elements of the total electronic angular momentum appear
(orbital plus spin).

The potential energy matrix $W_{kl}(x)$ is defined by (\ref{Wdef}) and
it is still rotationally invariant as shown by (\ref{Wklrotns}),
exactly as in the electrostatic model.  The only difference is that
$W_{kl}$ now contains contributions to the energy from the fine
structure.  Likewise, the molecular Hamiltonian in the
Born-Oppenheimer representation, $H_{{\rm mol},kl}$, is
(\ref{Hmolkldef}), the same as in the electrostatic model, as is the
projected Hamiltonian $K_{{\rm mol},kl}$.  

As for the angular momentum, it is more interesting to work with
$\Jvec=\Lvec+\Svec$ than with $\Lvec=\Lvec_n+\Lvec_e$ alone.  We
follow the steps of (\ref{Ldistribute}) in converting $\Jvec$ to the
Born-Oppenheimer representation,
\begin{eqnarray}
  \Jvec\Psi(\Xvec,\rvec,m) &=& (\Lvec_n + \Lvec_e+\Svec)
  \sum_k \psi_k(\Xvec)\,\phi_k(\Xvec;\rvec,m)
  =\sum_k[(\Lvec_n+\Lvec_e)\psi_k(\Xvec)]\,\phi_k(\Xvec;\rvec,m)
  \nonumber\\
  &&\qquad+\sum_k\psi_k(\Xvec)\,[(\Lvec_n+\Lvec_e+\Svec)\phi_k(\Xvec;\rvec,m)],
  \label{LdistributeFSeven}
\end{eqnarray}
where the differential operator $\Lvec_n+\Lvec_e$ is distributed as
before, while the operator $\Svec$ only acts on the second factor
(which depends on the spin quantum numbers $m$).   But by
(\ref{JtotbasiszeroFSeven}) the second major sum vanishes, as does the
term involving $\Lvec_e$ in the first sum.  The result is
\begin{equation}
  \Jvec=\Lvec+\Svec=\Lvec_n+\Lvec_e+\Svec
  \longleftrightarrow \Lvec_n\,\delta_{kl}.
  \label{JvecBOrepnFSeven}
\end{equation}
Thus, in the fine structure model with $N_e={\rm even}$, in the
Born-Oppenheimer representation, the nominal, orbital angular momentum
of the nuclei alone represents physically the total angular momentum
of the molecule, both nuclear and electronic, including the electron
spin. This is exact.

Thus conservation of angular momentum is represented in the
Born-Oppenheimer representation by (\ref{HmolLncomrel}), exactly as in
the electrostatic model.  That is, the components of $H_{{\rm
    mol},kl}$ must be scalars under nuclear orbital rotations.  That
they are follows from the transformation property of the derivative
couplings, (\ref{Fvecconjrel}), which is the same as in the
electrostatic model.

Finally, consider a solution $\psi_k$ for $k\in\Levels$ of the
Born-Oppenheimer version of the Schr\"odinger equation,
(\ref{BOScheqnmultilevel}), that is also an
eigenfunction of the matrix operators $L_n^2\,\delta_{kl}$ and
$L_{nz}\,\delta_{kl}$ with quantum numbers $(l,m_l)$.   Every
component $k\in\Levels$ of such a solution is an eigenfunction of $L_n^2$
and $L_{nz}$ with the same quantum numbers.  When this is
converted to a molecular wave function by (\ref{Psiintermsofpsi}) it
is automatically an eigenfunction of $J^2$ and $J_z$ with the same
quantum numbers, and this is exact.  

\subsection{Fine Structure Details, Odd Number of Electrons}
\label{FSdetailsodd}

\subsubsection{Basis States}
\label{FSdetailsbasis}

In the fine structure model with $N_e={\rm odd}$ the energy levels are
Kramers doublets.  See Sec.~\ref{oddnumber} for terminology regarding
``levels'' and ``surfaces.''  We define a subset of strongly coupled
surfaces
\begin{equation}
   \Levels=\{k_0,k_0+1,\ldots,k_0+N_s-1\},
   \label{AdefFSodd}
\end{equation}
where $N_s$ is the number of surfaces, which replaces (\ref{Adef}).
Now the number of levels is $N_l=2N_s$.  We denote the adiabatic basis
vectors by $\ket{ax;k\mu}$, $\mu=1,2$, which are energy eigenstates
for $k\in\Levels$,
\begin{equation}
  H_e(x) \,\ket{ax;k\mu} = \epsilon_k(x)\,\ket{ax;k\mu},
  \quad k\in\Levels,
  \label{adiabbasiskinAFSodd}
\end{equation}
where the energy depends on $k$ but not $\mu$ (this is the Kramers
degeneracy).  For $k\notin\Levels$ the vectors $\ket{ax;k\mu}$ form a
discrete, orthonormal basis that spans $\Sspace^\perp(x)$.  Because
$H_e(x)$ commutes with $T$, these basis vectors can be chosen to be
quaternionic, as we assume (see (\ref{Tbasistau})).  

Initially we make some assignment of these vectors along a section
$S$, that is, of phase and frame conventions so that the basis is
quaternionic.  This assignment cannot be smooth when $S$ contains a
degeneracy, that is, a crossing of two or more surfaces or Kramers
doublets.  The codimension of such degeneracies is different from the
case of the electrostatic model (generically 5 or sometimes 3 instead
of 2, see \cite{Mead80a, Mead87, MatsikaYarkony01,
  MatsikaYarkony02b}), but the fact remains that in general a
continuous assignment of adiabatic basis states on $S$ is impossible.
We accept the discontinuities and extend the definitions of the basis
vectors along rotational fibers by means of a modified rule, see
(\ref{basisrotxfmFSodd}) below, which differs from the ones
(\ref{basisrotxfm}) or (\ref{basisrotxfmFSeven}) used previously.
This rule guarantees that the adiabatic basis, so extended, remains
quaternionic.

Given the adiabatic basis there are various algorithms for defining a
diabatic basis, which is free of the singularities of the adiabatic
basis.  We denote the diabatic basis by $\ket{dx;k\mu}$.  If the
adiabatic basis is quaternionic, we must ask whether the diabatic
basis so constructed is too.  The answer depends on the algorithm, but
we have checked both the singular-value diabatic basis, which is due
to \cite{Pacheretal88, Pacheretal93}, and the parallel-transported
diabatic basis.  These bases were the subject of a recent study of
ours (\cite{LittlejohnRawlinsonSubotnik22}).  It turns out that if the
adiabatic basis is quaternionic, then the diabatic basis, constructed
by either of these two algorithms, is also quaternionic.  In this way
we can construct a diabatic basis on $S$ that is quaternionic; this
can then be propagated along rotation fibers by
(\ref{basisrotxfmFSodd}), giving us a smooth, quaternionic, diabatic
basis in a region of full dimensionality.

The two bases are connected by a unitary transformation,
\begin{equation}
  \ket{dx;k\mu}=\sum_{l,\nu}\ket{ax;l\nu}\,V_{l\nu,k\mu}(x)
  \label{ADTdefFSodd}
\end{equation}
the analog of (\ref{ADTdef}) in the electrostatic model.  Since the
two bases are quaternionic, the matrix $V_{l\nu,k\mu}$ is both unitary
and quaternionic, that is, the minor, $2\times 2$ matrices $V_{lk}$,
whose $(\nu\mu)$ components are $V_{l\nu,k\mu}$, are quaternions.  See
Appendix~\ref{quaternions}.  The $N_s \times N_s$ block of this matrix
of quaternions corresponding to the strongly coupled subspace belongs
to the unitary, quaternionic group $U(N_s,\Quaternions)$.  Unlike the
electrostatic case, the matrix $V_{l\nu,k\mu}$ is not constant along
rotational fibers, but rather satisfies
\begin{equation}
  V_{kl}(Rx_0)=u\,V_{kl}(x_0)\,u^{-1},
  \label{VklxfmrotnsFSodd}
\end{equation}
where $R=R(u)$.  This is written in terms of the minor matrices or
quaternions that make up $V$.

\subsubsection{Working Basis and Its Properties}
\label{FSdetailworkingbasis}

Thus we obtain a working basis, which we denote by simply
$\ket{x;k\mu}$, which is either the adiabatic basis $\ket{ax;k\mu}$
when that is smooth or else the diabatic basis $\ket{dx;k\mu}$.  The
vectors of the working basis are propagated along rotational fibers by
the rule,
\begin{equation}
  \ket{Rx_0;k\mu} = \sum_\nu U_e(u)\,\ket{x_0;k\nu}\,
  \bigl(u^{-1}\bigr)_{\nu\mu}.
  \label{basisrotxfmFSodd}
\end{equation}
which gives the basis vectors at $x=Rx_0$ in terms of those at $x_0$.
In this equation, $u$ on the right hand side means one of the two
elements of $SU(2)$ that satisfies $R=R(u)$, which differ by a sign.
Because there is an odd number of factors of $u$ contained in the
operator $U_e(u)$ and an extra one in the factor of $u^{-1}$, the
total number is even and the right hand side does not depend on which
of the two $u$'s is chosen.  This was one reason for introducing the
factor of $u^{-1}$ on the right hand side; without it, the formula
would not define a single-valued basis set along a rotational fiber.
This factor evidently causes a mixing among the Kramers pair as the
molecule is rotated.

The rule (\ref{basisrotxfmFSodd}) has several important properties.
The first is that if $\ket{x_0;k\mu}$ is quaternionic at $x_0$, then
$\ket{Rx_0;k\mu}$ is quaternionic at $x=Rx_0$.  The second is that if
$\ket{x_0;k\mu}$ is an electronic eigenstate at $x_0$ (which is the
case for $k\in\Levels$ in the adiabatic basis), then $\ket{Rx_0;k\mu}$ is
an electronic eigenstate at $x=Rx_0$, with the same eigenvalue.  The
proofs are given in Appendix~\ref{proofs}.

Another important property is the transformation law,
\begin{equation}
  \ket{Rx;k\mu} = \sum_\nu U_e(u)\,\ket{x;k\nu} \, 
  \bigl(u^{-1}\bigr)_{\nu\mu}
  \label{wbasisxfmrotsFSodd}
\end{equation}
which is like (\ref{basisrotxfmFSodd}) but with $x_0$ replaced by $x$.
Compare (\ref{basisrotxfm}) and (\ref{wbasisxfmrots}) in the
electrostatic model, and see the discussion below
(\ref{wbasisxfmrots}).  In particular, notice that
(\ref{basisrotxfmFSodd}) defines the basis states along a rotation
fiber, and (\ref{wbasisxfmrotsFSodd}) is a property of those basis
states, once defined.  The proof of (\ref{wbasisxfmrotsFSodd}) is
given in Appendix~\ref{proofs}.

If we had chosen $u$ instead of $u^{-1}$ in (\ref{basisrotxfmFSodd})
then we would have a single-valued definition of phase and frame
conventions along a rotational fiber, but (\ref{wbasisxfmrotsFSodd})
would not be valid, with either $u$ or $u^{-1}$.  This was the main
reason we chose $u^{-1}$ in (\ref{basisrotxfmFSodd}), which we believe
is the most satisfactory definition of phase and frame conventions
along rotational fibers in the fine structure model with $N_e={\rm
  odd}$.  This choice leads to a simple interpretation of the wave
function $\psi_{k\mu}(\Xvec)$ for fixed $k$ and $\mu=1,2$ as belonging
to a particle of pseudo-spin $1/2$, moving on a multidimensional,
potential energy surface.

Given our basis states $\ket{x;k\mu}$ we define basis wave functions
$\phi_{k\mu}(\Xvec;\rvec,m)$ by (\ref{phiketT=-1}), that is, with the
double index $(k\mu)$.  The transformation to the Born-Oppenheimer
representation is given by
\begin{equation}
   \Psi(\Xvec,\rvec,m)=\sum_{k\mu} \psi_{k\mu}(\Xvec)\,
	\phi_{k\mu}(\Xvec;\rvec,m),
   \label{PsipsiphiFS-1allk}
\end{equation}
which we can also write as
\begin{equation}
    \Psi(\Xvec,\rvec,m) \longleftrightarrow \psi_{k\mu}(\Xvec).
    \label{PsiarrowpsiFSodd}
\end{equation}
Equation~(\ref{PsipsiphiFS-1allk}) involves a sum on both $k$ and
$\mu$ and is the exact representation of the wave functions
$\Psi(\Xvec,\rvec,m)$ (and is not to be confused with
(\ref{PsipsiphiFS-1}) which applies to a single-surface problem).
Double indices also appear in the Born-Oppenheimer representation of
operators,
\begin{equation}
    A \longleftrightarrow A_{k\mu,l\nu},
    \label{AarrowAkmulnuFSodd}
\end{equation}
where the right hand side can also be written in terms of minor
matrices $A_{kl}$.

\subsubsection{Representation of Angular Momentum}

We now let $R$ in (\ref{wbasisxfmrotsFSodd}) be infinitesimal, and
proceed as in the derivation of (\ref{Ltotbasiszero}) or
(\ref{JtotbasiszeroFSeven}).   We invoke the infinitesimal version
of $u^{-1}$,
\begin{equation}
	u^{-1} = 1+\frac{i}{2}\,\theta\nvechat\cdot\sigmavec,
	\label{u-1infinitesimal}
\end{equation}
and follow the steps leading to (\ref{Ltotbasiszero}), finding
\begin{equation}
    \Jvec\ket{x;k\mu}=
    (\Lvec_n+\Lvec_e+\Svec)\ket{x;k\mu}=
	\frac{\hbar}{2}\sum_\nu
	\ket{x;k\nu}\,(\sigmavec)_{\nu\mu},
    \label{JtotbasisKFSodd}
\end{equation}
which takes the place of (\ref{Ltotbasiszero}) in the electrostatic
model or (\ref{JtotbasiszeroFSeven}) in the fine structure model with
$N_e={\rm even}$.  The nonzero result on the right hand side comes
from the factor of $u^{-1}$ that was inserted into the transformation
law (\ref{basisrotxfmFSodd}).

This allows us to find the total angular momentum of the molecule in
the Born-Oppenheimer representation.  We proceed as in
(\ref{LdistributeFSeven}) using the expansion (\ref{PsipsiphiFS-1allk}),
finding 
\begin{eqnarray}
    \Jvec\Psi(\Xvec,\rvec,m) &=& (\Lvec_n+\Lvec_e+\Svec)
    \sum_{k\mu} \psi_{k\mu}(\Xvec)\,
    \phi_{k\mu}(\Xvec;\rvec,m)
    \nonumber\\
    &=& \sum_{k\mu}[(\Lvec_n+\Lvec_e)\psi_{k\mu}(\Xvec)]\,
     \phi_{k\mu}(\Xvec;\rvec,m)
    \nonumber\\
    && +\sum_{k\mu}\psi_{k\mu}(\Xvec)\,
    [(\Lvec_n+\Lvec_e+\Svec)\phi_{k\mu}(\Xvec;\rvec,m)].
\end{eqnarray}
In the first major sum on the right the contribution from $\Lvec_e$
vanishes as before but now in view of (\ref{JtotbasisKFSodd}) the
second major sum is nonzero.  Altogether we find
\begin{equation}
    \Jvec\Psi(\Xvec,\rvec,m) = 
    \sum_{k\mu}[\Lvec_n\psi_{k\mu}(\Xvec)]\,
	\phi_{k\mu}(\Xvec;\rvec,m)
    +\sum_{k\mu\nu}\psi_{k\mu}(\Xvec) \,
    \phi_{k\nu}(\Xvec;\rvec,m)\left(
    \frac{\hbar}{2}\sigmavec\right)_{\nu\mu}.
\end{equation}
Swapping $\mu$ and $\nu$ in the second term makes both sums a linear
combination of $\phi_{k\mu}(\Xvec;\rvec,m)$, so that the result can be
written,
\begin{equation}
    \Jvec \longleftrightarrow \Lvec_n \,\delta_{kl} \,\delta_{\mu\nu}
    +\delta_{kl}\left(\frac{\hbar}{2}\sigmavec\right)_{\mu\nu},
\end{equation}
or, in terms of minor matrices,
\begin{equation}
   \Jvec \longleftrightarrow \delta_{kl}\, (\Lvec_n + \Kvec).
   \label{JvecarrowIvecFSodd}
\end{equation}
Here $\Kvec=(\hbar/2)\sigmavec$ is a vector of minor matrices which
act on a Born-Oppenheimer wave function $\psi_{k\mu}$ just by matrix
multiplication in the pseudo-spin index $\mu$. 

This motivates the definition $\Ivec=\Lvec_n+\Kvec$ made earlier (see
(\ref{Ivecdef})).  The angular momentum $\Ivec$ is associated with
rotation operators that are parameterized by $u\in SU(2)$ and that we
denote by $U_i(u)$.  They are defined by
\begin{equation}
    U_i(u) = U_i(\nvechat,\theta)
    =\exp\left(-\frac{i}{\hbar}\,\theta\nvechat\cdot\Ivec\right),
    \label{Uidef}
\end{equation}
and their action on wave functions $\psi_\mu(\Xvec)$ is given by
\begin{equation}
    \bigl(U_i(u)\psi\bigr)_\mu(\Xvec) =
	\sum_\nu u_{\mu\nu}\, \psi_\nu\bigl(R^{-1}\Xvec\bigr).
    \label{Uiaction}
\end{equation}
This implies the representation property, $U_i(u_1)U_i(u_2)
=U_i(u_1u_2)$.  In (\ref{Uiaction}) we write $\psi_\mu(\Xvec)$ without
a $k$-index because it is useful to think of this as the wave function
of a pseudoparticle of spin $1/2$.   The Born-Oppenheimer wave
function $\psi_{k\mu}(\Xvec)$ can be thought of as an
infinite-dimensional vector of such wave functions, indexed by $k$.

\subsubsection{Derivative Couplings}
\label{derivcouplingsFSodd}

The fine structure derivative couplings when $N_e={\rm odd}$ have been
defined in (\ref{FvecdefT=-1}).   They are denoted
$\Fvec_{\alpha;k\mu,l\nu}(x)$, or, as minor matrices which turn out to
be quaternions, as $\Fvec_{\alpha;kl}(x)$.   These form an
anti-Hermitian matrix of quaternions, as noted in
(\ref{FisantiHermT=-1v3}).

These transform along rotational fibers according to
\begin{equation}
    \Fvec_{\alpha;kl}(Rx) = u[R\Fvec_{\alpha;kl}(x)]u^{-1},
    \label{FvecxfmrotsFSodd}
\end{equation}
where $R=R(u)$.  This may be compared to (\ref{Fvecxfmrots}), which
applies both in the electrostatic model and in the fine structure
model with $N_e={\rm even}$.  The derivation is similar if slightly
more complicated.  

We also require the transformation of the derivative couplings under
conjugation by $U_i(u)$.   A purely spatial vector like $\Xvec_\alpha$
transforms as a vector operator,
\begin{equation}
   U_i(u)\,\Xvec_\alpha\,U_i(u)^\dagger = R^{-1}\Xvec_\alpha,
   \label{XvecUiconj}
\end{equation}
since the pseudo-spin part of $U_i(u)$ does nothing, while in the case
of a minor matrix $\omega$ with no spatial dependence we have
\begin{equation}
   U_i(u) \, \omega U_i(u)^\dagger = u\omega u^{-1},
   \label{qUiconj}
\end{equation}
since the spatial part does nothing.  Therefore
\begin{equation}
    U_i(u) \, \Fvec_{\alpha;kl}(x) \, U_i(u)^\dagger
    = u\Fvec_{\alpha;kl}\bigl(R^{-1}x\bigr)u^{-1}
    = R^{-1}\Fvec_{\alpha;kl}(x),
    \label{FvecconjrelFSodd}
\end{equation}
where in the last step we use a version of (\ref{FvecxfmrotsFSodd})
with $R\to R^{-1}$ and $u\to u^{-1}$.  This may be compared to
(\ref{Fvecconjrel}) which applies both in the electrostatic model and
in the fine structure model with $N_e={\rm even}$.  

Finally, we compute the components of the derivative couplings in a
purely rotational direction, defining $\delta x$ and
$\delta\Xvec_\alpha$ as in (\ref{infinitesR}) and (\ref{deltaxdef}).
Then we find
\begin{equation}
   \sum_{\alpha=1}^{N-1} \delta\Xvec_\alpha\cdot
   \Fvec_{\alpha;k\mu,l\nu}(x) = \frac{i}{\hbar}\,\theta\nvechat\cdot
   \left[-\matrixelement{x;k\mu}{(\Lvec_e+\Svec)}{x;l\nu}
   +\frac{\hbar}{2}\,\delta_{kl}\,\sigmavec_{\mu\nu}\right],
   \label{FrotnlcompFSodd}
\end{equation}
where we use (\ref{JtotbasisKFSodd}).  This may be compared to
(\ref{Frotnlcomp}) in the electrostatic model or
(\ref{FrotnlcompFSeven}) in the fine structure model with $N_e={\rm
even}$.

\subsubsection{The Hamiltonian}
\label{HamFSodd}

When the electronic Hamiltonian $H_e(x)$ is converted to the
Born-Oppenheimer representation, it becomes a matrix,
\begin{equation}
   W_{k\mu,l\nu}(x) =
   \matrixelement{x;k\mu}{H_e(x)}{x;l\nu},
   \label{Wkmulnudef}
\end{equation}
which can be interpreted in terms of minor matrices denoted
$W_{kl}(x)$.   Since $H_e(x)$ commutes with time reversal, these minor
matrices are quaternions; and since $H_e(x)$ is Hermitian, these
quaternions satisfy $W_{kl}(x) = \overline{W_{lk}(x)}$.  It then
follows from (\ref{FSHeconjR}) and (\ref{wbasisxfmrotsFSodd}) that
\begin{equation}
    W_{kl}(Rx) = u W_{kl}(x) u^{-1}.
    \label{WklrotnsFSodd}
\end{equation}
In the case of the fine structure model with $N_e={\rm odd}$, the
matrix $W_{kl}(x)$ is not constant along rotational fibers.  An
exception is the diagonal elements; these are real quaternions,
$W_{kk}=\overline{W_{kk}}$, that is, as minor matrices they are a
multiple of the identity, so the factors of $u$ and $u^{-1}$ in
(\ref{WklrotnsFSodd}) cancel.   In particular, for a single surface
problem $W_{k_0k_0}(x)$ is the Kramers degenerate eigenvalue
$\epsilon_{k_0}(x)$. 

To transform the molecular Hamiltonian to the Born-Oppenheimer
representation we start with the momentum, which transforms according
to 
\begin{equation}
    \Pvec_\alpha \longleftrightarrow \Pvec_\alpha\,\delta_{kl}
    \,\delta_{\mu\nu} - i\hbar \Fvec_{\alpha;k\mu,l\nu}(x).
    \label{PvecalphaBOrepnFSodd}
\end{equation}
This simplifies in the language of minor matrices,
\begin{equation}
    \Pvec_\alpha \longleftrightarrow \Pvec_\alpha\,\delta_{kl}
    -i\hbar \Fvec_{\alpha;kl}(x),
    \label{PvecalphaBOrepnFSoddv2}
\end{equation}
where now it is understood that $\Pvec_\alpha$ is multiplied by the
unit minor matrix.   The result (\ref{PvecalphaBOrepnFSoddv2}) looks
exactly the same as (\ref{PalphaBO}) in the electrostatic model except
that now the operator in the Born-Oppenheimer representation is
interpreted as a minor matrix.  

Similarly, the molecular Hamiltonian $H_{\rm mol}$ becomes a matrix of
minor matrices $H_{{\rm mol},kl}$ in the Born-Oppenheimer
representation, the formula for which is (\ref{Hmolkldef}), exactly as
in the electrostatic model but now reinterpreted as a relation among
minor matrices.  Of course, one must respect the order of
multiplication of minor matrices when expanding the products
shown.  

Since in the molecular representation $H_{\rm mol}$ commutes with
$\Jvec$, in the Born-Oppenheimer representation we expect the matrix
$H_{{\rm mol},kl}$ of minor matrices to commute with
$\Ivec\,\delta_{kl}=(\Lvec_n+\Kvec)\delta_{kl}$, another such matrix.
See (\ref{JvecarrowIvecFSodd}).  But since the latter matrix is a
multiple of the identity $\delta_{kl}$, we expect
\begin{equation}
    [\Ivec,H_{{\rm mol},kl}]=[\Lvec_n+\Kvec,H_{{\rm mol},kl}]=0,
    \label{HmolLncomrelFSodd}
\end{equation}
which takes the place of (\ref{HmolLncomrel}) in the electrostatic
model or the fine structure model with an even number of electrons.
That is, there is now a contribution $\Kvec$ to the angular momentum,
and everything is interpreted as minor matrices.  

Equation~(\ref{HmolLncomrelFSodd}) holds if and only if every
component $H_{{\rm mol},kl}$ of the Hamiltonian commutes with
$U_i(u)$, defined by (\ref{Uidef}) or (\ref{Uiaction}), that is, if
every such component transforms as a scalar under conjugation by
$U_i(u)$.   To show that they do we start with the fact that
$\Pvec_\alpha$ transforms as a vector operator, just like
$\Xvec_\alpha$ (see (\ref{XvecUiconj})), and so does
$\Fvec_{\alpha;kl}$ (see (\ref{FvecconjrelFSodd})).  Therefore dot
products that look like $\Pvec\cdot\Pvec$, $\Pvec\cdot\Fvec$ or
$\Fvec\cdot\Fvec$ are scalars.   As for the potential energy matrix
$W_{kl}(x)$, we have
\begin{equation}
   U_i(u) \, W_{kl}(x) \, U_i(u)^\dagger
   =u\,W_{kl}\bigl(R^{-1}x\bigr) \, u^{-1}
   = W_{kl}(x),
   \label{U_iconjWklFSodd}
\end{equation}
where in the last step we use (\ref{WklrotnsFSodd}) with
$R$ and $u$ swapped with $R^{-1}$ and $u^{-1}$.  Thus we check
(\ref{HmolLncomrelFSodd}).  

In the Born-Oppenheimer approximation $H_{{\rm mol},kl}$ is replaced
by its projected version $K_{{\rm mol},kl}$, which is the same
when $k,l\in\Levels$ and zero otherwise.   Therefore, just as in the other
models, $K_{{\rm mol},kl}$ commutes with $\Ivec$ since $H_{{\rm
mol},kl}$ does.   The Born-Oppenheimer approximation to the
Schr\"odinger equation can be written exactly as in
(\ref{BOScheqnmultilevel}), except that now $K_{{\rm mol},kl}$ is
a minor matrix of operators and $\psi_k$ must be understood as a
2-component pseudo-spinor with components $\psi_{k\mu}$, $\mu=1,2$.  A
solution of this equation will also be an eigenfunction of
$I^2\,\delta_{kl}$ and $I_z\,\delta_{kl}$ with quantum numbers
$(i,m_i)$ if each spinor component $\psi_k$ is an eigenfunction of
$I^2$ and $I_z$ with the same quantum numbers.  Such an eigenfunction,
when converted to the molecular representation via
(\ref{PsipsiphiFS-1allk}), will automatically be an eigenfunction of
$J^2$ and $J_z$ with the same quantum numbers $(i,m_i)$.

\section{The Dressing Transformation}
\label{dressingxfm}

As explained, the Born-Oppenheimer approximation or its generalization
to multisurface problems can be described as just throwing away off
block-diagonal elements of $H_{{\rm mol},kl}$, that is, for
$k\in\Levels$ and $l\notin\Levels$ or $k\notin\Levels$ and
$l\in\Levels$.  A more satisfactory procedure, however, is to remove
these off-diagonal terms by means of unitary transformations.  This is
conveniently done in the Born-Oppenheimer representation by mapping
operators into their Weyl transforms (\cite{McDonald88}), and using a
version of the Moyal bracket (\cite{Moyal49}) for carrying out the
perturbation expansion.  The main ideas of this approach are given by
\cite{LittlejohnFlynn91}, and applied to the Born-Oppenheimer
approximation by \cite{WeigertLittlejohn93}. See also
\cite{Panatietal02, Teufel03}.  In this section we shall briefly
summarize the ideas and conclusions, enough to show their relevance to
the subject of angular momentum.  For simplicity we shall describe the
situation in the electrostatic model.

The method generates a power series in $\kappa^2$, where
$\kappa=(m/M)^{1/4}$ is the usual Born-Oppenheimer ordering parameter.
When we refer to ``first order,'' we shall mean, first order in
$\kappa^2$, while ``second order'' means order $\kappa^4$, etc.  

In the first step we transform the molecular Hamiltonian,
\begin{equation}
   \Hbar_{{\rm mol},1} = U_1\,H_{\rm mol}\, U_1^\dagger,
   \label{Kmoldef}
\end{equation}
where $U_1$ is a unitary transformation that is designed make the
off-diagonal terms of $\Hbar_{{\rm mol},1}$ vanish to lowest order in
$\kappa^2$.  All operators are expressed in the Born-Oppenheimer
representation, that is, as matrices (thus, for example, $\Hbar_{{\rm
mol},1kl}$ and $U_{1,kl}$), but the subscripts are suppressed in
(\ref{Kmoldef}).   The unitary operator $U_1$ is expressed in terms of
an anti-Hermitian generator $G_1$, $U_1=\exp(G_1)$, so that, for
any operator $A$, we have
\begin{equation}
    U_1 A U_1^\dagger = A + [G_1,A] + \frac{1}{2!}
	[G_1,[G_1,A]] + \ldots
    \label{dressingexpansion}
\end{equation}
This series of iterated commutators turns into a power series in
$\kappa^2$.  

A single unitary transformation of the type shown in (\ref{Kmoldef})
is capable of removing the off-diagonal terms only to first order, but
there will remain second-order terms.  We can apply a second unitary
transformation to remove these, leaving behind third-order,
off-diagonal terms.  Thus to fully remove these terms we must
contemplate an infinite number of unitary transformations, for which
we write, $U=\ldots U_3U_2U_1$, where each $U_n$ has a generator
$G_n$.  The generators $G_n$ turn out to be of order $\kappa^{2n}$.
In practice, the first generator $G_1$ is the most important, and is
responsible for most of what is described in the literature as
``nonadiabatic corrections'' to the Born-Oppenheimer approximation.

It turns out that $G_1$ contains energy denominators of the form
$\epsilon_k(x)-\epsilon_l(x)$, where $k\in\Levels$ and
$l\notin\Levels$.  Thus when this energy difference is of order
$\kappa^2$ or smaller, the expansion (\ref{dressingexpansion}) breaks
down and levels cannot be separated by adiabatic means.  This gives
some quantitative meaning to the notion of ``strongly coupled''
levels, which were discussed in Sec.~\ref{coupledss}.

This sequence of unitary transformations produces Hamiltonians
$\Hbar_{{\rm mol},1}$, $\Hbar_{{\rm mol},2}$, etc.  We will write
simply $\Hbar_{\rm mol}$ for $\Hbar_{{\rm mol},\infty}$, so that
$\Hbar_{\rm mol} = U H_{\rm mol} U^\dagger$.   We will refer to $U$ as
the ``dressing transformation'' and $\Hbar_{\rm mol}$ as the ``dressed
Hamiltonian.''  The latter is block-diagonal to all orders in
$\kappa^2$.  

We can now distinguish what we will call the ``original Born-Oppenheimer
representation,'' what was called in Sec.~\ref{molecBOreps} simply the
``Born-Oppenheimer representation,'' from the ``dressed Born-Oppenheimer
representation.''  There is, of course, also the molecular
representation, which was described in Sec.~\ref{molecBOreps}.
Physical observables have different operators representing them in the
different representations.   For example, the physical observables
which are the Jacobi vectors
are represented by the operators $\Xvec_\alpha$ in the molecular
representation, that is, the operators are multiplication by
$\Xvec_\alpha$.   As described in Sec.~\ref{molecBOreps}, these
physical observables are represented by the matrices
$\Xvec_\alpha\,\delta_{kl}$ in the original Born-Oppenheimer
representation, which we can write simply as $\Xvec_\alpha$ if we
remember that an identity matrix is implied.  In the dressed
Born-Oppenheimer representation, however, they are represented by the
operators $\Xvecbar_\alpha = U\,\Xvec_\alpha \,U^\dagger$, which are
not the same as $\Xvec_\alpha$.   In fact, to first order in
$\kappa^2$, we have
\begin{equation}
    \Xvecbar_\alpha = \Xvec_\alpha + [G_1,\Xvec_\alpha] + \ldots.
    \label{dressXalpha}
\end{equation}
Similar statements can be made about the nuclear momenta, which
in the original Born-Oppenheimer representation involve the derivative
couplings (see (\ref{PalphaBO})).  

As for the total orbital angular momentum of the molecule, we have
seen that it is represented by $\Lvec=\Lvec_n+\Lvec_e$ in the
molecular representation and $\Lvec_n\,\delta_{kl}$ in the original
Born-Oppenheimer representation, which we can abbreviate as simply
$\Lvec_n$ if we remember that it is multiplied by the identity
matrix.  As for the dressed Born-Oppenheimer representation, the same
physical observable is represented by
\begin{equation}
   U\,\Lvec_n \,U^\dagger = U \sum_{\alpha=1}^{N-1} \Xvec_\alpha
   \times \Pvec_\alpha \, U^\dagger
   =\sum_{\alpha=1}^{N-1} \Xvecbar_\alpha \times 
   \Pvecbar_\alpha.
   \label{Lvecdressed}
\end{equation}
But the dressing of $\Lvec_n$ involves a series of commutators with
the generators $G_n$, such as shown in (\ref{dressingexpansion}).  The
generators $G_n$ are responsible for transforming the rotationally
invariant Hamiltonian $H_{\rm mol}$ to its diagonalized version,
$\Hbar_{\rm mol}$, which is also rotationally invariant.   The
generators $G_n$ that do this are themselves rotationally invariant, so they
commute with angular momentum and all the correction terms in power
series like (\ref{dressingexpansion}) vanish.  Thus we have
\begin{equation}
    \sum_{\alpha=1}^{N-1} \Xvec_\alpha \times
    \Pvec_\alpha = \sum_{\alpha=1}^{N-1}
    \Xvecbar_\alpha \times \Pvecbar_\alpha,
\end{equation}
to all orders of the Born-Oppenheimer expansion.  The dressing does
nothing to the angular momentum $\Lvec_n$, which represents physically
the total angular momentum of the molecule in both the original
Born-Oppenheimer representation and the dressed version of it.

In the case of single-surface problems the dressing transformation
creates a $1\times1$ block $\Hbar_{{\rm mol},k_0k_0}$ on the diagonal,
that is decoupled from all other levels to all orders of $\kappa$.
Thus in the dressed Born-Oppenheimer representation the solution of
the Schr\"odinger equation is a simple product form as seen in
(\ref{Psipsiphi}), to all orders of $\kappa^2$.  For this reason we
suspect that there is a connection between Moyal perturbation theory,
as discussed here, and the method of ``exact factorization''
(\cite{AbediMaitraGross10, AbediMaitraGross12, Cederbaum13,
Scherreretal15, SchildAgostiniGross16, Requistetal16,
MartinazzoBurghardt22}).  The possibility of such a connection is a
project for the future.

The dressed Hamiltonian $\Hbar_{\rm mol}$ may be compared to $K_{\rm
  mol}$, which was obtained in Sec.~\ref{HmolBOrepn} by throwing away
off-diagonal terms.  Both are block-diagonal, but they are not the
same Hamiltonians.  This is because the dressing transformation
modifies the diagonal block, adding extra terms to it.  The first such
term appears at second order.  This term has evidently been discovered
independently several times (\cite{Moodyetal89, WeigertLittlejohn93,
  Goldhaber05}), but it has had no impact on the chemical literature.
It is of order $\kappa^4$ and is therefore small, but it is of the
same order as terms that are routinely discussed in connection with
Born-Oppenheimer theory.  We will say more about this term in future
publications.

\section{Discussion and Conclusions}
\label{conclusions}

We have presented an in-depth analysis of angular momentum in the
Born-Oppenheimer theory of polyatomic molecules, revealing exact
equivalences among its various representations.  We have done this
both in the electrostatic model and when fine structure effects and
electron spin are included.  Several new results are reported
regarding the transformation laws under rotations of the Hamiltonian,
basis states and derivative couplings.  A dressing
transformation that replaces the Born-Oppenheimer approximation
reveals further exact equivalences among representations of angular
momentum. 

Finally, note that we have not made any semiclassical approximations
above, and the exact equivalences described above hold rigorously.
That being said, the results do have clear implications for
semiclassical calculations.  In particular, within surface hopping
calculations (\cite{FatehiAlguireShaoSubotnik11}), there has been a
long literature regarding questions of how to treat electronic
momentum and how to conserve momentum with electron translation
factors (\cite{BatesMcCarroll58, SchneidermanRussek69, Delos81,
IllescasRiera98}).  These questions arise because the electronic
momentum is hidden in the phase conventions of the Born-Oppenheimer
representation.  To that extent this article has pointed out that
similar questions can also be raised in the context of rotations and
angular momentum.  This line of study will be pursued in a subsequent
publication as well.

\begin{acknowledgments}
  RL and JR would like to thank the CHAMPS project (Chemistry and
  Mathematics in Phase Space) and Bristol University for their support
  during the workshop, ``Exact Factorization, Geometric Phase and
  Bohmian Mechanics,'' September 19--20, 2022, which helped advance
  the ideas presented in this article.  JES was supported by the
  National Science Foundation under Grant No.  CHE-2102402.
\end{acknowledgments}

\appendix
\section{Time Reversal}
\label{timerev}

Time reversal is covered in texts (\cite{Messiah66,
  SakuraiNapolitano11}) and specifically in molecular theory
(\cite{Mead79, Mead80a, Rosch83, Mead87, KoizumiSugano95,
  JohnssonAitchison97, SchonKoppel98, MatsikaYarkony01,
  MatsikaYarkony02a, MatsikaYarkony02b}).  We prefer an approach based
on invariant subspaces.

There are at least two distinct time reversal operators relevant to
this article: the one that acts on electronic wave functions
$\phi(\rvec)$, relevant in the electrostatic model, and the one that
acts on wave functions $\phi(\rvec,m)$, relevant in the fine structure
model.  In the electrostatic model time reversal acts on electronic
wave functions according to
\begin{equation}
	(T\phi)(\rvec)=\phi(\rvec)^*,
	\label{TdefES}
\end{equation}
that is, by simple complex conjugation, so $T^2=+1$.  In the fine
structure model the wave function $\phi(\rvec,m)$ depends on spin and
time reversal acts according to
\begin{equation}
  (T\phi)(\rvec,m) = \sum_{m'_1} \ldots \sum_{m'_{N_e}}
  \tau_{m_1m'_1} \ldots \tau_{m_{N_e}m'_{N_e}}\,
  \phi(\rvec,m')^*,
  \label{TdefFS}
\end{equation}
where $m'$ is a primed version of (\ref{mdef}), where each magnetic
quantum number $m'_i$, $i=1,\ldots,N_e$ ranges over $\pm1/2$, and where
$\tau$ is the matrix
\begin{equation}
  \tau=e^{-i\pi\sigma_y/2} =
  \left(\begin{array}{cc}
      0 & -1 \\
      1 & 0
      \end{array}\right).
    \label{taudef}
\end{equation}
The matrix $\tau$ is the spin rotation $u(\yvechat,\pi)$ in the
notation (\ref{udef}).  We abbreviate an equation like (\ref{TdefFS})
by writing
\begin{equation}
  (T\phi)(\rvec,m) = \sum_{m'} (\tau \ldots \tau)_{mm'}\,
  \phi(\rvec,m')^*.
  \label{TdefFSabbrev}
\end{equation}

\subsection{Properties of $T$}
\label{propertiesofT}

Time reversal $T$ is an antiunitary operator, $T^\dagger T = T
T^\dagger =1$, that satisfies $T^2=+1$ in the electrostatic model or
in the fine structure model with $N_e={\rm even}$, and $T^2=-1$ in the
fine structure model with $N_e={\rm odd}$.  These are the only
properties of $T$ that are needed for the rest of this appendix.

If $\Hspace$ is a Hilbert space upon which $T$ acts and $\Sspace
\subset \Hspace$ is a subspace, then we say that $\Sspace$ is {\it
  invariant} under $T$ if for every $\ket{\psi}\in\Sspace$,
$T\ket{\psi} \in \Sspace$ (that is, $T$ maps $\Sspace$ into itself).
Important examples of invariant subspaces include the entire Hilbert
space ($\Sspace=\Hspace$) and eigenspaces (possibly degenerate) of a
Hamiltonian that commutes with $T$.  Many other examples appear in
this article.  As for energy eigenspaces, note that if
$H\ket{\psi}=E\ket{\psi}$, $T^\dagger H T=H$, and
$\ket{\phi}=T\ket{\psi}$, then $H\ket{\phi}=E\ket{\phi}$.  That is,
$T$ maps energy eigenstates into other energy eigenstates of the same
energy.  This does not say whether $\ket{\phi}$ is linearly
independent of $\ket{\psi}$.

Antiunitary operators map scalar products into their complex
conjugates, that is, if $\ket{\psi'}=T\ket{\psi}$ and
$\ket{\phi'}=T\ket{\phi}$, then $\braket{\phi'}{\psi'} =
\braket{\phi}{\psi}^*=\braket{\psi}{\phi}$.  This in turn means that
time reversal maps orthonormal frames into other orthonormal frames,
that is, if $\{\ket{n},n=1,\ldots,N\}$ is an orthonormal frame,
$\braket{n}{m}=\delta_{nm}$, and if $\ket{n'}=T\ket{n}$, then
$\braket{n'}{m'}=\delta_{nm}$.  The frame $\{\ket{n}\}$ need not be
complete (it need not span the whole Hilbert space).  This does not
say whether the new frame $\{\ket{n'}\}$ is linearly independent of
the old one $\{\ket{n}\}$.

In the following we will take subspaces that are invariant under $T$
and break them down into smaller, mutually orthogonal subspaces that
are also invariant under $T$.  First we note that if $\Sspace$ is
invariant under $T$, then it is also invariant under $T^\dagger$, as
follows from the fact that $T^\dagger=\pm T$ (the same sign as in
$T^2=\pm1$).   

Now let $\Sspace \subset \Hspace$ be an invariant subspace under $T$,
let $\Aspace \subset \Sspace$ be a subspace of $\Sspace$ that is
also invariant under $T$, and let $\Bspace \subset \Sspace$ be the
space orthogonal to $\Aspace$ inside $\Sspace$, so that
\begin{equation}
  \Sspace=\Aspace \oplus \Bspace.
  \label{SABspaces}
\end{equation}
Then $\Bspace$ is invariant under $T$.  To prove this we note that a
vector $\ket{\phi}\in \Bspace$ if and only if $\ket{\phi}\in\Sspace$
and $\braket{\phi}{\psi}=0$ for all $\ket{\psi}\in\Aspace$.  Now let
$\ket{\phi}\in\Bspace\subset\Sspace$ and let
$\ket{\phi'}=T\ket{\phi}$, so that $\ket{\phi'}\in\Sspace$.  We wish
to show that $\ket{\phi'}$ is orthogonal to all
$\ket{\psi}\in\Aspace$, hence $\ket{\phi'}\in\Bspace$.  First we note
that $\braket{\phi'}{\psi} = (\bra{\phi}T^\dagger)\ket{\psi}
=[\bra{\phi}(T^\dagger\ket{\psi})]^* = \braket{\phi}{\psi'}^*$, where
$\ket{\psi'}=T^\dagger\ket{\psi}$.  But $\ket{\psi'}\in\Aspace$ since
$\Aspace$ is invariant under $T^\dagger$, and therefore the scalar
product vanishes.  Therefore $\ket{\phi'}\in\Bspace$, and $\Bspace$ is
invariant under $T$.

\subsection{Case $T^2=+1$}
\label{Tcase+1}

Now we specialize to the case $T^2=+1$.  Let $\Sspace$ be a subspace
invariant under $T$, with $\dim\Sspace\ge1$.  Then $\Sspace$
possesses a 1-dimensional, invariant subspace.  To prove this let
$\ket{\psi}\in\Sspace$ be a nonzero vector and consider the two
vectors $\ket{\psi}$ and $T\ket{\psi}$.   If these are linearly
dependent, then $\ket{\psi}$ spans a 1-dimensional, invariant subspace
of $\Sspace$.   If they are linearly independent then
$\ket{\psi}+T\ket{\psi}$ is nonzero and spans a 1-dimensional,
invariant subspace.  

Then the space inside $\Sspace$, complementary and orthogonal to this
1-dimensional, invariant subspace, is also invariant under $T$, so, if
its dimensionality is $\ge1$, it also possesses an invariant,
1-dimensional subspace.  Proceeding by induction, we see that if
$\Sspace$ is finite-dimensional, then it can be decomposed into a set
of mutually orthogonal, 1-dimensional subspaces, each invariant under
$T$.  We will assume that the same holds when $\Sspace$ is
infinite-dimensional.

Now let $\ket{e}$ be a unit vector inside a 1-dimensional, invariant
subspace.  Then $T\ket{e} = e^{i\alpha}\ket{e}$ for some phase factor
$e^{i\alpha}$, since both $\ket{e}$ and $T\ket{e}$ are bases inside
the 1-dimensional subspace.  Then defining
$\ket{\phi}=e^{i\alpha/2}\ket{\psi}$, we have 
\begin{equation}
  T\ket{\phi} = e^{-i\alpha/2}\,T\ket{\psi}=\ket{\phi},
\end{equation}
and $\ket{\phi}$ is invariant under $T$.  That is, by a phase
convention we can make the basis in a 1-dimensional, invariant
subspace invariant under time reversal.  This applies to each of the
subspaces into which $\Sspace$ of the previous paragraph was
decomposed, so we see that in the case $T^2=+1$, a $T$-invariant
subspace $\Sspace$ always possesses a $T$-invariant orthonormal basis.

In particular, bases can be chosen inside the eigenspaces of a
Hamiltonian that commutes with $T$ that are $T$-invariant, that is,
such a Hamiltonian always possesses a $T$-invariant energy eigenbasis.

It is easy to show that the matrix elements of a $T$-invariant
operator such as the Hamiltonian in a $T$-invariant basis are real.

Now let $\Sspace\subset\Hspace$ be a subspace of a Hilbert space and
let $\{\ket{e_n}\}$ and $\{\ket{f_n}\}$ be two orthonormal bases in
$\Sspace$.  Then these bases are connected by a unitary 
transformation,
\begin{equation}
  \ket{f_n} = \sum_m \ket{e_m}\,U_{mn},
  \label{fnxfm}
\end{equation}
where $U^\dagger U= UU^\dagger=1$.  If however $\Sspace$ is
$T$-invariant, as are the two bases, $T\ket{e_n}=\ket{e_n}$,
$\ket{f_n}=T\ket{f_n}$, then it is easy to show that
$U_{mn}=U^*_{mn}$,  that is, $U$ is a real, orthogonal matrix.  If
$N=\dim\Sspace$, then $U\in O(N)$, the latter being the group that
connects choices of $T$-invariant, orthonormal bases in the case
$T^2=+1$.  Conversely, if the basis $\ket{e_n}$ in (\ref{fnxfm}) is
$T$-invariant and if $U$ is real orthogonal, then the basis
$\ket{f_n}$ is also $T$-invariant.

As a special case, if $N=1$, the group $O(1)$ consists of just two
matrices $(+1)$ and $(-1)$, so the choice of a $T$-invariant basis
reduces to the choice of a $\pm$ sign.

\subsection{Case $T^2=-1$}
\label{Tcase-1}

Now let $T^2=-1$ and, as before, let $\Sspace$ be a subspace invariant
under $T$ such that $\dim\Sspace\ge1$.  Then $\Sspace$ does not
possess any 1-dimensional, invariant subspaces but it does possess a
2-dimensional invariant subspace.  To prove this let $\ket{\psi}\ne0$
be a vector in $\Sspace$ and notice that $\ket{\phi}=T\ket{\psi}$ is
also a nonzero vector in $\Sspace$ since $\Sspace$ is $T$-invariant
and $T$ preserves norms.  These vectors are orthogonal,
\begin{equation}
  \braket{\phi}{\psi} = (\bra{\psi}T^\dagger)\ket{\psi}
  =[\bra{\psi}(T^\dagger\ket{\psi})]^*
  =-[\bra{\psi}(T\ket{\psi})]^*
  =-\braket{\psi}{\phi}^*=-\braket{\phi}{\psi}=0,
\end{equation}
where we use $T^\dagger=-T$, and therefore linearly independent.  They
span a 2-dimensional, invariant subspace of $\Sspace$, since
\begin{equation}
  T(a\ket{\psi}+b\ket{\phi})=a^*\ket{\phi}-b^*\ket{\psi},
\end{equation}
where we use $T\ket{\phi} = T^2\ket{\psi}=-\ket{\psi}$.   

Thus $\dim\Sspace\ge2$ and $\Sspace$ possesses a 2-dimensional,
invariant subspace.  If $\Sspace$ is the eigenspace of a $T$-invariant
Hamiltonian, then this implies that all eigenvalues are at least
2-fold degenerate (the usual statement of Kramers degeneracy).  But
this means that the space inside $\Sspace$ that is orthogonal and
complementary to this 2-dimensional, invariant subspace is also
invariant, so, by induction, we can continue to split off
2-dimensional, invariant subspaces until $\Sspace$ is exhausted (if
ever).   If $\Sspace$ is finite-dimensional, this implies that
$\Sspace$ can be decomposed into a set of mutually orthogonal,
2-dimensional, invariant subspaces; and we will assume that this can
also be done when $\Sspace$ is infinite-dimensional.  If $\Sspace$ is
finite-dimensional, then $\dim\Sspace=2N$ is even.

\subsubsection{Quaternionic bases}
\label{qbases}  

If the vector $\ket{\psi}$ of the preceding paragraphs is a unit
vector then we may call it $\ket{1}$; and then
$\ket{\phi}=T\ket{\psi}=T\ket{1}$ is also a unit vector, call it
$\ket{2}$.   Then the set $\{\ket{1},\ket{2}\}$ forms an orthonormal
basis in the invariant subspace that they span, such  that
$T\ket{1}=\ket{2}$ and $T\ket{2}=-\ket{1}$.   Doing the same for each
of the 2-dimensional, invariant subspaces into which an invariant
subspace $\Sspace$ is decomposed, we obtain an orthonormal basis
inside $\Sspace$, $\{\ket{k\mu},k=1,\ldots,N,\mu=1,2\}$, such that
\begin{equation}
  T\ket{k\mu} = \sum_\nu \ket{k\nu}\,\tau_{\nu\mu},
  \label{Tbasistau}
\end{equation}
where $\tau$ is given by (\ref{taudef}) and where $\dim\Sspace=2N$.
We shall call such a basis {\it quaternionic}.  (In equations like
this we label the rows and columns of $\tau$ by $1,2$; in other places
by $1/2,-1/2$.)

Now let $\ket{e;k\mu}$ and $\ket{f;l\nu}$ be two quaternionic bases on
a $T$-invariant subspace $\Sspace$.   (The symbols $e$ and $f$ just
distinguish  the two bases.)  Since the bases are orthonormal,
they must be connected by a unitary matrix,
\begin{equation}
  \ket{f;l\nu}=\sum_{k\mu}\ket{e;k\mu}\,U_{k\mu,l\nu},
  \label{feconnect}
\end{equation}
where 
\begin{equation}
  \sum_{n\sigma}U_{k\mu,n\sigma}\,U^*_{l\nu,n\sigma}=
  \delta_{kl}\,\delta_{\mu\nu}=
  \sum_{n\sigma}U^*_{n\sigma,k\mu}\,U_{n\sigma,l\nu}.
  \label{Uisunitary}
\end{equation}
In cases like this we shall view the matrix $U_{k\mu,l\nu}$ as a
``major'' matrix that is composed of $2\times2$ blocks that we will
call ``minor'' matrices.  If we write simply $U_{kl}$, we shall mean
the minor matrix whose $(\mu\nu)$ component is $U_{k\mu,l\nu}$.   With
this understanding, (\ref{Uisunitary}) can be written,
\begin{equation}
  \sum_n U_{kn}(U_{ln})^\dagger = \delta_{kl}
  =\sum_n (U_{nk})^\dagger\,U_{nl},
  \label{Uunitaryminor}
\end{equation}
where we use parentheses to make it clear, for example, that
$(U_{nk})^\dagger$ is the Hermitian conjugate of the minor matrix
$U_{nk}$, and where $\delta_{kl}$ is understood to be multiplied by
the identity minor matrix.

Now applying $T$ to both sides of (\ref{feconnect}) we obtain
\begin{equation}
  \sum_{\nu'}\ket{f;l\nu'}\,\tau_{\nu'\nu}=
  \sum_{k\mu\mu'}\ket{e;k\mu'}\,\tau_{\mu'\mu}\,U^*_{k\mu,l\nu}.
\end{equation}
We multiply this by $(\tau^\dagger)_{\nu\sigma}$ and sum over $\nu$,
to obtain
\begin{equation}
  \ket{f;l\sigma}=\sum_{k\mu\mu'\nu}\ket{e;k\mu'}\,\tau_{\mu'\mu}
  \,U^*_{k\mu,l\nu}\,(\tau^\dagger)_{\nu\sigma}
  =\sum_{k\mu'} \ket{e;k\mu'} \,
  \bigl(\tau U^*_{kl}\tau^\dagger\bigr)_{\mu'\sigma}
  =\sum_{k\mu'}\ket{e;k\mu'}\,U_{k\mu',l\sigma},
\end{equation}
where in the last step we have used (\ref{feconnect}) again.  Then,
since the vectors $\ket{e;k\mu'}$ are linearly independent, we obtain
$U_{kl} = \tau U^*_{kl}\tau^\dagger$, an equation connecting minor
matrices.   Multiplying this on the left by $\tau^\dagger$ and on the
right by $\tau$, we obtain
\begin{equation}
  \tau^\dagger U_{kl} \tau = U^*_{kl},
\end{equation}
showing that the minor matrices of $U$ are quaternions (see
(\ref{qcrit})).   

Thus, $U$ belongs to the group $U(N,\Quaternions)$, the set of
$N\times N$ unitary matrices of quaternions.  These matrices satisfy
\begin{equation}
  \sum_n U_{kn}\, \overline{U_{ln}} = \delta_{kl}
  =\sum_n \overline{U_{nk}}\,U_{nl},
\end{equation}
which is (\ref{Uunitaryminor}) written in quaternionic language.
Conversely we can show that if the basis $\ket{e;k\mu}$ is
quaternionic and $U\in U(N,\Quaternions)$, then the basis
$\ket{f;l\nu}$ defined by (\ref{feconnect}) is also quaternionic.   

In the special case $N=1$, which applies to a single surface problem
with an odd number of electrons (a single Kramers doublet), the group
$U(1,\Quaternions)$ consists of unit quaternions, those for which
$U_{11} \overline{U_{11}}=1$.  As noted in Appendix~\ref{quaternions}
this is the group $SU(2)$.  Its role in this context was appreciated
by \cite{Mead87}.

\subsubsection{Quaternionic matrix elements}
\label{quatmes}

Finally, let $A$ be a linear operator that commutes with time reversal,
$T^\dagger A T=A$.  Then the matrix elements of $A$ in a quaternionic
basis form minor matrices that are quaternions.  To prove this we
consider the matrix elements of $A$ with respect to a quaternionic
basis $\{\ket{k\mu}\}$,
\begin{eqnarray}
  A_{k\mu,l\nu} &=& \matrixelement{k\mu}{A}{l\nu} =
  \bra{k\mu}(T^\dagger A T\ket{l\nu}) =
  [(\bra{k\mu}T^\dagger) A (T\ket{l\nu})]^*
  \nonumber\\
  &=& \left[\sum_{\mu'\nu'}(\bra{k\mu'}\tau^*_{\mu'\mu})
  A(\ket{l\nu'}\tau_{\nu'\nu})\right]^*
  =\left[\sum_{\mu'\nu'} (\tau^\dagger)_{\mu\mu'} \,
  A_{k\mu',l\nu'}\, \tau_{\nu'\nu}\right]^*,
  \label{Aisquatproof}
\end{eqnarray}
or, in terms of minor matrices, $A_{kl} = (\tau^\dagger A_{kl}
\tau)^*$.   Now taking the complex conjugate of both sides and
comparing to (\ref{qcrit}) we see that $A_{kl}$ is a quaternion.

\section{Quaternions}
\label{quaternions}

It is well known that the quantum mechanics of systems with an odd
number of fermions is conveniently described in terms of quaternions
(\cite{Dyson62, Finkelstein62, Rosch83, Avronetal88,
  JohnssonAitchison97, Zhang97, SaueAaJensen99, DeLeoScolarici00,
  SadovskiiZhilinskii22}).  Quaternions also play an important role in
representation theory (\cite{Simon96}).  In this appendix we summarize
what is needed for this article.  Our treatment is similar to that of
\textcite{Rosch83}.

For the purposes of this article a quaternion is a $2\times2$ matrix
of the form
\begin{equation}
	q=a-i\bvec\cdot\sigmavec,
	\label{qdef}
	\end{equation}
where $a$ and $\bvec=(b_1,b_2,b_3)$ are real.  We denote the set of
quaternions by $\Quaternions$.  Hamilton's unit quaternions $\iquat$,
$\jquat$, $\kquat$ are identified with the matrices $-i\sigma_i$,
$i=1,2,3$.  By this definition the matrices (\ref{qdef}) form a
representation of the algebra of quaternions (matrix multiplication
and inversion are equivalent to the same operations on quaternions,
etc).

The quaternion conjugate to $q$, denoted ${\bar q}$, is
obtained from $q$ by the replacement  $\bvec\to-\bvec$.  Interpreted as a
matrix, this is the same as forming the Hermitian conjugate; therefore
we will write ${\bar q} = q^\dagger$, and note that
$\overline{q_1q_2}=\overline{q_2}\,\overline{q_1}$.  A quaternion $q$ is
said to be {\em real} if $\bvec=0$, that is, ${\bar q}=q$.  The square
magnitude of a quaternion is
\begin{equation}
	|q|^2={\bar q}q=q{\bar q}=a^2+b_1^2+b_2^2+b_3^2 = \det q.
	\label{qsqared}
	\end{equation}
As for complex conjugation, by $q^*$ we mean the complex conjugate of
the $2\times2$ matrix (\ref{qdef}).

A $2\times2$ matrix $q$ is a quaternion, according to (\ref{qdef}), if
and only if
\begin{equation}
	\tau^\dagger q \tau = q^*,
	\label{qcrit}
	\end{equation}
where $\tau$ is given by (\ref{taudef}).  Notice that $\tau$ is the
basis quaternion $\jquat$. 

A unit quaternion $q$ is one for which $|q|^2=1$.  The set of unit
quaternions, interpreted as matrices, is the group $SU(2)$.  An
arbitrary quaternion can be written as $q=\rho u$, where $\rho\ge 0$
is real and $u$ is an element of $SU(2)$, which is unique if $\rho>0$.

\section{Rotation Groups $SO(3)$ and $SU(2)$}
\label{rotations}

We let $R\in SO(3)$ be a proper rotation, which we parameterize in
axis-angle form, $R=R(\nvechat,\theta)$, where the unit vector
$\nvechat$ is the axis of the rotation and $\theta$ is the angle.  All
proper rotations are covered if $\nvechat$ runs over the unit sphere
and $0\le\theta\le\pi$.  If $0<\theta<\pi$ the axis-angle
representation is unique, but if $\theta=0$ then $R(\nvechat,0)=I$
(the identity) for all axes $\nvechat$, and if $\theta=\pi$ then
$R(\nvechat,\pi)=R(-\nvechat,\pi)$.  This shows that the space of
proper rotations, the group manifold $SO(3)$, is diffeomorphic to the
real projective space $\Reals P^3$.  (Two spaces are diffeomorphic if
their points can be placed in a one-to-one correspondence in a smooth
manner.  It means that the spaces are identical from a differentiable
or topological standpoint.  The space $\Reals P^3$ is the 3-sphere
$S^3$ with antipodal points identified.)

The group $SU(2)$ consists of matrices that can be parameterized in
axis-angle form,
\begin{equation}
  u(\nvechat,\theta)=e^{-i\theta\nvechat\cdot\sigmavec/2}
    =\cos(\theta/2) -i\nvechat\cdot\sigmavec\sin(\theta/2),
  \label{udef}
\end{equation}
where $\sigmavec$ is the vector of Pauli matrices.  All of $SU(2)$ is
covered if the axis $\nvechat$ runs over the unit sphere and
$0\le\theta\le2\pi$.  The representation is unique if $0<\theta<2\pi$
but when $\theta=0$, $u(\nvechat,0)=1$ for all $\nvechat$, and when
$\theta=2\pi$, $u(\nvechat,2\pi)=-1$ for all $\nvechat$.  This means
that the group manifold $SU(2)$ is diffeomorphic to the 3-sphere
$S^3$.

The projection from $SU(2)$ to $SO(3)$ is given by
\begin{equation}
  R_{ij} = \frac{1}{2}\tr(u^\dagger \sigma_i u \sigma_j).
  \label{R(u)def}
\end{equation}
One can show that the matrix $R$ defined by this equation belongs to
$SO(3)$ if $u \in SU(2)$, so it defines a map or function $:SU(2)\to
SO(3)$.  We will denote the function by $R(u)$; it is a group
homomorphism,
\begin{equation}
  R(u_1)R(u_2) = R(u_1u_2),
  \label{R(u)homomorph}
\end{equation}
and thus $SO(3)$ forms a representation of $SU(2)$.  The map
(\ref{R(u)homomorph}) is two-to-one, since $R(u)=R(-u)$.  The map
preserves the axis and angle, that is,
$R\bigl(u(\nvechat,\theta)\bigr) = R(\nvechat,\theta)$.

Sometimes it is desirable to invert (\ref{R(u)def}), that is, given
$R\in SO(3)$ we wish to find $u$.  The answer can be given by using
the axis-angle parameterization; if we write $R=R(\nvechat,\theta)$,
then the two elements of $SU(2)$ that satisfy (\ref{R(u)def}) are
$\pm u(\nvechat,\theta)$.

\section{Some Proofs}
\label{proofs}

We prove the statements made below (\ref{basisrotxfmFSodd}), which
concern the consequences of that formula.  First, suppose a basis is
quaternionic at $x_0$,
\begin{equation}
  T\ket{x_0;k\mu}=\sum_\nu \ket{x_0;k\nu}\,\tau_{\nu\mu},
  \label{Tonbasisatx0}
\end{equation}
see (\ref{Tbasistau}), and suppose that $\ket{Rx_0;k\mu}$ is given by
(\ref{basisrotxfmFSodd}).  Then we have
\begin{eqnarray}
  T\ket{Rx_0;k\mu} &=& \sum_\nu U_e(u)\bigl(T\ket{x_0;k\nu}\bigr)
  \bigl(u^{-1})^*_{\nu\mu}=\sum_{\nu\sigma}
  U_e(u)\,\ket{x_0;k\sigma}\,\tau_{\sigma\nu}\,\bigl(u^{-1})^*_{\nu\mu}
  \nonumber\\
  &=&\sum_\sigma U_e(u)\,\ket{x_0;k\sigma}\,
  \bigl(\tau u^{-1*}\bigr)_{\sigma\mu},
  \label{Tketstage1}
\end{eqnarray}
where in the first step we use the fact that $T$ commutes with
rotations.  But since $u^{-1}\in SU(2)$ it is a quaternion and
satisfies $u^{-1*} = \tau^\dagger\, u^{-1}\,\tau$, see (\ref{qcrit}).
Thus $\tau\,u^{-1*} = u^{-1}\,\tau$, and (\ref{Tketstage1}) becomes
\begin{equation}
  \sum_{\nu\sigma}
  U_e(u)\,\ket{x_0;k\sigma}\,\bigl(u^{-1})_{\sigma\nu}\,
  \tau_{\nu\mu}
  =\sum_\nu \ket{Rx_0;k\nu}\,\tau_{\nu\mu}.
\end{equation}
Thus, the basis $\ket{Rx_0;k\mu}$ at the rotated point $x=Rx_0$ is
also quaternionic.

Next, suppose a basis vector at $x_0$ is an energy eigenvector,
\begin{equation}
  H_e(x_0) \,\ket{x_0;k\mu} = \epsilon_k(x_0)\,\ket{x_0;k\mu},
\end{equation}
where the energy does not depend on $\mu$ as indicated.  Then
\begin{eqnarray}
  H_e(Rx_0)\,\ket{Rx_0;k\mu} &=& U_e(u) H_e(x_0) U_e(u)^\dagger
  \sum_\nu U_e(u) \,\ket{x_0;k\nu}\,\bigl(u^{-1}\bigr)_{\nu\mu}
  \nonumber\\
  &=&\epsilon_k(x_0)\,\sum_\nu U_e(u)\,\ket{x_0;k\nu}\,
  \bigl(u^{-1}\bigr)_{\nu\mu} 
  =\epsilon_k(x_0) \,\ket{Rx_0;k\mu},
\end{eqnarray}
where in the first step we use (\ref{FSHeconjR}) and
(\ref{basisrotxfmFSodd}).  Thus, the rule (\ref{basisrotxfmFSodd})
maps energy eigenbases at $x_0$ into  those at $x=Rx_0$, without
changing the eigenvalues.

To prove (\ref{wbasisxfmrotsFSodd}) we let $R_1,R_2\in SO(3)$,
corresponding to $u_1,u_2 \in SU(2)$, and we write $x_1=R_1x_0$ and
$x_2=R_2x_1$.  Then we have
\begin{eqnarray}
  \ket{x_2;k\mu} &=& \ket{R_2x_1;k\mu} = \ket{R_2R_1x_0;k\mu}
  =\sum_\nu U_e(u_2u_1)\,\ket{x_0;k\nu} \, (u_2u_1)^{-1}_{\nu\mu}
  \nonumber\\
  &=&\sum_{\nu\sigma} U_e(u_2)\,U_e(u_1)\,\ket{x_0;k\nu}\,
  \bigl(u_1^{-1}\bigr)_{\nu\sigma} \,
  \bigl(u_2^{-1}\bigr)_{\sigma\mu}
  \nonumber\\
  &=&\sum_\sigma U_e(u_2)\,\ket{x_1;k\sigma}\,
  \bigl(u^{-1}_2\bigr)_{\sigma\mu}.
\end{eqnarray}
Now making the replacements $x_1\to x$, $R_2\to R$ and $u_2\to u$, we
obtain (\ref{wbasisxfmrotsFSodd}).   The proof would not work if we had
used $u$ instead of $u^{-1}$ in (\ref{wbasisxfmrotsFSodd}).

\bibliography{angmom.bib}

\end{document}